\documentclass[letterpaper,11pt]{article}
\usepackage{jheppub, tikz, physics, enumitem, tikz-feynman, indentfirst, color, bbold, comment, parskip}

\usepackage[mathscr]{eucal}

\usepackage[linkcolor=blue, colorlinks=true, citecolor=blue, urlcolor=blue]{hyperref}

\newcommand{\n}{d}

\newcommand{\0}{{}^{(0)}}
\newcommand{\1}{{}^{(1)}}
\newcommand{\hmn}{h_{\mu\nu}}
\newcommand{\g}{\lambda}

\newcommand{\be}{\begin{equation}}
\newcommand{\ee}{\end{equation}}

\tikzfeynmanset{compat=1.1.0}
\usetikzlibrary{intersections, shapes.arrows, shapes.misc, decorations.pathreplacing, decorations.pathmorphing, angles, quotes, calc, patterns, arrows.meta, fadings, tikzmark}
\tikzset{cross/.style={cross out, draw=black, minimum size=2*(#1-\pgflinewidth), inner sep=0pt, outer sep=0pt}, cross/.default={1pt}}

\title{\boldmath Gravitational Corrections to OPE Coefficients}

\author[a]{Ameya Chavda,}
\author[b]{Alessandro Podo,}
\author[a, c]{John Staunton}

\affiliation[a]{Center for Theoretical Physics and Department of Physics, Columbia University, \\
New York, NY 10027, USA}

\affiliation[b]{Institut des Hautes \'Etudes Scientifiques,\\
91440 Bures-sur-Yvette, France}

\affiliation[c]{Center for Particle Cosmology, Department of Physics and Astronomy,\\ University of Pennsylvania, 209 South 33rd St, Philadelphia, PA 19104, USA}

\emailAdd{ameya.chavda@columbia.edu}
\emailAdd{podo@ihes.fr}
\emailAdd{jstaunto@sas.upenn.edu}

\abstract{The short-distance singular behavior of correlation functions of quantum field theories in curved spacetime is a universal property shared by all well-behaved states (known as Hadamard states).
For two-point functions this takes the form of the Hadamard tail of free scalars and its generalizations to spinning fields and interacting theories. In this work, we compute similar universal curvature corrections to curved-space OPE coefficients in a short-distance expansion for interacting quantum field theories in general dimensions. We work at leading order in the curvature but for arbitrary backgrounds. Examples include weakly coupled scalar field theories with cubic and quartic interactions as well as strongly coupled conformal field theories.  The gravitationally dressed OPE coefficients allow us to test explicitly, and confirm, the validity of the smearing-in-time procedure discussed by Witten in the construction of the algebra of observables near an observer's worldline in curved spacetime.}

\begin{document}
\maketitle
\flushbottom

\section{Introduction}
\label{sec:intro}

Quantum field theory (QFT) in curved spacetime is a rich and important subject, with applications ranging from early universe cosmology to the physics of black holes. Yet, few general results on the dynamics of quantum fields in an arbitrary curved background are available. The goal of this work is to apply the formalism presented in a companion paper, Ref.~\cite{Chavda:2025ijz}, to the computation of gravitational corrections to the operator product expansion (OPE) of interacting QFTs in curved spacetime. The method leverages the equivalence principle, together with a derivative expansion, to treat an arbitrary curved spacetime as a weak perturbation of flat space, and thus apply flat-space perturbation theory. Let us summarize the logic of this approach.

General Relativity (GR) can be understood as the effective field theory for a massless spin two particle, the graviton. From this point of view, the Einstein--Hilbert Lagrangian and the cosmological constant terms are only the first two in an infinite series of terms in the action. Higher derivative corrections are suppressed by the short distance cutoff, $\ell_{\text{UV}}$, which could be of order of the Planck scale $\ell_{\text{UV}} \sim 1/M_{\text{Pl}}$. Even if the short distance quantum gravity dynamics is unknown, we can still compute, within the effective field theory (EFT) regime of validity, low energy quantum gravity effects associated to a non-analytic dependence on $\ell_{\text{UV}}$ and hence momentum, e.g. as induced by the leading Einstein--Hilbert term. This is because the UV dependence is captured by local terms, suppressed by higher integer powers of $\ell_{\text{UV}}$, or equivalently momentum. 

Consider now a situation in which the curvature of the background is of the order of some given distance scale, $L$. That is: the Ricci scalar, $R$, is of order $L^{-2}$ and higher derivative (and curvature) terms are controlled by higher powers of $1/L$. For most situations of physical interest, there is a separation of scales between $L$ and $\ell_{\text{UV}}$, and we can consider the former as an IR scale, $\ell_{\text{IR}} \equiv L$. Then, if we are interested in a range of scales, $s$, such that
\begin{equation}
\ell_{\text{UV}} \ll s \ll \ell_{\text{IR}},
\end{equation}
we can consider an expansion in $s/\ell_{\text{IR}}$ and compute corrections suppressed by this ratio in an EFT approach. The corrections that have a non-analytic dependence on $s/\ell_{\text{IR}}$ will be genuine \textit{short-distance} predictions of QFT in curved spacetime. Here, by short distance, we mean $s \ll \ell_{\text{IR}}$, while we are setting $\ell_{\text{UV}} \rightarrow 0$ for simplicity. In other words, in this work, we will always neglect $\ell_{\text{UV}}$ suppressed quantum gravity corrections, and focus on short-distance non-analytic effects at scales $s \ll \ell_{\text{IR}}$.

The simplest example of such a short-distance curvature correction that can be computed in EFT is in the two-point function of two free scalar fields in $d$ spacetime dimensions~\cite{Chavda:2025ijz}
\begin{equation}
\label{eq:2ptHadamard}
\expval{T\phi\left(x_1\right)\phi\left(x_2\right)} = \frac{C_d}{\left(s^2 + i\varepsilon\right)^{d/2 - 1}}\left(\mathcal{D}^{1/2}\left(x_1, x_2\right) + \frac{1-6\xi}{12\left(d - 4\right)} s^2\, R\left(x_0\right)\right),
\end{equation}
where $s$ is the geodesic distance between points $x_1$ and $x_2$ and $\mathcal{D}$ is the van Vleck determinant that measures the extent to which geodesics focus or defocus between $x_1$ and $x_2$. Here $\xi$ is the non-minimal gravitational coupling, and $C_d$ fixes the normalization of the two-point correlator. In the limit $d\to 4$, the pole in the second term gives rise to a logarithmic singularity whose coefficient depends on the curvature of the background, called the \textit{Hadamard tail}, which vanishes only for conformal coupling. Generalizations to spin-1 and spin-2 fields, and to two-point functions in interacting CFTs have been discussed in Ref.~\cite{Chavda:2025ijz}.

\textbf{Summary of Results.} Our goal in this paper will be to generalize Eq.~\eqref{eq:2ptHadamard} to higher point correlation functions, whose information is encoded in an OPE coefficient. First, in Section~\ref{s:intscalars}, we extract the gravitational corrections to leading and subleading OPE coefficients for the merging of two scalars in theories involving a cubic and quartic interaction. As an application, we will compute the OPE coefficients for a quartic theory at the Wilson--Fischer fixed point. We will then apply this strategy to general conformal field theories (CFTs) in Section~\ref{s:Conformal Primaries}, once again computing the leading and subleading OPE coefficients for the merging of two conformal scalar primaries. Finally, in Section~\ref{s:algebra}, with explicit results in hand, we will test the temporal smearing procedure of Witten~\cite{Witten:2023qsv} to construct the algebra of observables in a local region near an observer's worldline.

Our results rest on a number of assumptions. First, we will assume an external non-dynamical background, meaning we neglect any back-reaction. Second, we will always compute correlation functions in \textit{Hadamard states}, i.e. states which reduce to the Poincar\'{e} vacuum at short enough distances. Most states of interest are Hadamard, including the Bunch--Davies vacuum in de Sitter space. Finally, we will always define operators in flat space and then couple them to an external (background) gravitational metric. We therefore consider ordinary local QFTs in flat space and then couple them to gravity through their stress-energy tensor. 

\textbf{Comparison to Previous Literature.} The Hadamard form of two-point functions for free fields in arbitrary backgrounds has been studied by solving the Green's function equation on an arbitrary curved background, such as in Refs.~\cite{DeWitt:1960fc, Allen:1987bn, Parker:2009uva, Toms:2014tia, Frob:2017gez}. The results are sometimes expressed as recursion relations for the coefficients of different singular structures in the two-point function.

Similar universal gravitational corrections are expected for the OPE coefficients in interacting theories, but their analysis is complicated by the lack of a simple classical analog (such as the Green's function equation for the two-point function). One proposal was put forward by Hollands in Ref.~\cite{Hollands:2006ag}, which uses local Wick expansions of correlation functions to isolate the desired OPE coefficient and express them as integrals involving Hadamard distributions. Furthermore, the derivative of OPE coefficients with respect to the coupling parameters in a given interacting theory can be related to products of other OPE coefficients as shown in Refs.~\cite{Hollands:2017chb,Frob:2020kiz, Klehfoth:2022vgx}. For a review of these ``flow relations," see Ref.~\cite{Hollands:2023txn}. Other studies of interacting quantum field theories in an arbitrary curved spacetime include Refs.~\cite{Drummond:1977dg, Brown:1980qq, Hathrell:1981zb, Jack:1983sk}, whose focus was on perturbative contributions to the Weyl anomaly coefficients and on singularities in two-point functions. Finally, there is recent work studying the gravitational OPEs in CFTs, first in a conformally flat background and then extending it to non-conformally flat spacetimes with an ambient space formalism \cite{Konechny:2026bqg, Das:2026tjz}. 

In this paper, on the other hand, we continue to develop the complementary approach put forward in Ref.~\cite{Chavda:2025ijz}. In particular, the equivalence principle allows us to study the short-distance behavior of correlators by taking advantage of the fact that the spacetime is locally flat and curvature corrections can be represented as weak metric insertions in flat-space correlators. Then, using an operator product expansion, we find that the gravitationally corrected OPE coefficients are integrals over flat-space OPE coefficients. In this approach, we do not write a generally covariant answer until the end of the calculation, so that the requirement of general covariance does not complicate intermediate steps in the calculation. The use of diagrams to systematically incorporate gravitational effects has been used previously, see e.g. Refs.~\cite{Barvinsky:1993en,Godazgar:2018boc, Komissarov:2022gax}. 

As we will see, our approach uses a systematic expansion in terms of hard and soft momenta. This allows us to extract OPE coefficients by studying different regions in real or kinematic space. Our calculations are then made simpler by leveraging well-developed techniques for flat-space Feynman integrals, such as the method of regions \cite{Beneke:1997zp}. This is particularly useful in our analysis of CFT correlators, which can be carried on for generic background metrics that are not conformally flat. We will work at leading order in the curvature and derivative expansions, neglecting corrections of order $R^2$ or $\partial^2 R$. Higher-order corrections can be included systematically but their study is left for future investigations.

\textbf{Notation and Conventions.} Throughout this paper, we will work in $d > 2$ spacetime dimensions, unless otherwise specified, and use the mostly plus signature $\left(-, +, \dots, +\right)$. GR conventions will closely follow those of \cite{Wald:1984rg}. In particular, the Riemann tensor is
\begin{equation}
R_{\mu\nu\rho} {}^{\sigma} = \partial_\nu \Gamma^\sigma_{\hphantom{\sigma}\mu\rho} - \partial_\mu \Gamma^\sigma_{\hphantom{\sigma}\nu\rho} + \Gamma^\lambda_{\hphantom{\lambda}\mu\rho}\Gamma^\sigma_{\hphantom{\sigma}\lambda\nu} - \Gamma^\lambda_{\hphantom{\lambda}\nu\rho}\Gamma^\sigma_{\hphantom{\sigma}\lambda\mu}.
\end{equation}
Moreover, the Ricci tensor is $R_{\mu\nu} = R_{\mu\rho\nu}^{\hphantom{\mu\rho\nu}\rho}$ and the Ricci scalar is $R = g^{\mu\nu} R_{\mu\nu}$. Our Fourier transform conventions are as follows:
\begin{equation}
\tilde{f}(k) = \int \! f(x)\,  e^{-ik \cdot x}\, \dd[\n]{x} \qquad \text{and} \qquad f(x) = \int \! \tilde{f}(k) \, e^{ik\cdot x} \, \frac{\dd[\n]{k}}{\left(2\pi\right)^\n}.
\end{equation}
To simplify the notation we will drop the tilde from Fourier transformed quantities, relying on the argument (position or momenta) to distinguish between $f(x)$ and $f(k)\equiv \tilde{f}(k)$.
We will also often use the following abbreviation of momentum space integrals:
\begin{equation}
\int_{k_1, \dots,\, k_l} \equiv \int \frac{\dd[\n]{k_1}}{\left(2\pi\right)^\n} \dots \frac{\dd[\n]{k_l}}{\left(2\pi\right)^\n}.
\end{equation}
Finally, we will use  $x^2 \equiv \eta_{\mu\nu} x^\mu x^\nu$ for squared distances as measured by the flat-space metric, $s^2$ for squared geodesic distances as measured by the curved metric, $ds^2 = g_{\mu\nu} \dd{x}^\mu \dd{x}^\nu$, and we will raise and lower indices using $\eta_{\mu\nu}$ only. The flat-space metric in Euclidean signature is denoted $\delta_{\mu\nu}$.

\section{The general framework for correlation functions}
\label{s:general framework}

Before delving into concrete calculations, let us summarize the method used both in this paper and the companion paper Ref.~\cite{Chavda:2025ijz}. Consider a QFT with action $S = S[\phi, g]$ on a fixed, non-dynamical background. By the equivalence principle, we can choose locally inertial coordinates around any point $x_0$, $g_{\mu\nu} (x)= \eta_{\mu\nu} + h_{\mu\nu}(x)$ with $h_{\mu\nu}(x_0)=\partial_\rho h_{\mu\nu}(x_0)=0$ so that $h_{\mu \nu}$ is small at distances from $x_0$ much shorter than the curvature scale $\ell_{\rm IR}$. To first order in $h_{\mu\nu}$, 
\begin{equation}
S\left[\phi, g\right] = S_0\left[\phi, \eta\right] + \frac{1}{2}\int h_{\mu\nu}\left(x\right) T^{\mu\nu}\left(x\right) \dd[d]{x} + \dots,
\end{equation}
where the ellipsis contains higher order corrections and the stress-energy tensor is 
\begin{equation}
T_{\mu\nu}\left(x\right) \equiv -\frac{2}{\sqrt{-g}} \eval{\frac{\delta S\left[\phi, g\right]}{\delta g^{\mu\nu}}}_{g = \eta}. 
\end{equation}
The leading gravitational correction to any time-ordered correlator is then a flat-space correlator with a single stress tensor insertion, 
\begin{equation}
\label{eq:first_order}
\begin{aligned}
\expval{ T \big(O_1(x_1) O_2(x_2) \dots \big)}_g & \simeq \langle T \big(O_1(x_1) O_2(x_2) \dots \big) \rangle_{0} \\ 
& \quad + \frac{i}{2} \int \dd[d]{y}  \, \langle T \big(O_1(x_1) T^{\mu\nu} (y) O_2(x_2) \dots \big)\rangle_{0} \, h_{\mu\nu}(y) \;, \end{aligned}
\end{equation}
where $O$ is any operator. At leading order, it follows that the curvature correction to an $n$-point function requires only the knowledge of $(n+1)$-point correlators in flat space. 

The computation of these corrections to correlators can be used to determine corrections to their OPE coefficients. In flat space, the OPE of a scalar operator $\phi$ with itself reads
\begin{equation}
\label{eq:OPE}
\mathcal{\phi}\left(x\right)\mathcal{\phi}\left(0\right)\overset{x \to 0}{\simeq} \sum_{\ell}\sum_{O_{\ell}}C_{\phi\phi}^{O_{\ell}}\left(x^2\right)x^{\mu_1} \cdots x^{\mu_\ell} O_{\mu_1\cdots \mu_\ell}(0) + \dots,
\end{equation}
where the OPE coefficients $C_{\phi\phi}^{O_{\ell}}$ are generically singular for $x^2 \to 0$. In a curved background, it is convenient to resum the insertions of $\hmn$ into curved space OPE coefficients $\hat{C}$
\begin{equation}
\phi(x_1)\,\phi(x_2)\overset{x_{1,2}\to x_0}{\simeq}\sum_{\mathcal{O}}\hat C^{\mathcal{O}}_{\phi\phi}(x_1,x_2;x_0)\,\mathcal{O}(x_0)\, + \dots,
\label{eq:curved_OPE}
\end{equation}
which now depend on $x_0$ because the background generically breaks translation invariance. OPE coefficients can be extracted by considering correlation functions in appropriate coincident limits. Since the stress-energy tensor can be inside or outside the radius of convergence of any OPE we attempt to use, we are naturally led to break up the integral into regions. This will be explored in more detail in Section~\ref{s:Conformal Primaries}.

As an example, the coefficient of the identity, $\hat C^{\mathbb{1}}_{\phi\phi}$ is state independent and reproduces the Hadamard tail. In other words, for a free scalar field, using Eq.~\eqref{eq:2ptHadamard}:
\begin{equation}
\label{eq:2ptHadamard_OPE}
\hat{C}^{\mathbb{1}}_{\phi\phi} = \frac{C_d}{\left(s^2 + i\varepsilon\right)^{d/2 - 1}}\left(\mathcal{D}^{1/2}\left(x_1, x_2\right) + \frac{1-6\xi}{12\left(d - 4\right)} s^2\, R\left(x_0\right) \right).
\end{equation}
When interactions are included, there will be further contributions. This will be studied in the following sections. 

\section{Interacting Scalars in curved space}
\label{s:intscalars}

The simplest case we can study is that of an interacting massless scalar field, $\phi$, with a cubic interaction. In the OPE language, that corresponds to the coefficient $\hat{C}_{\phi\phi}^\phi$:
\begin{equation}
\mathcal{\phi}\left(x_1\right)\mathcal{\phi}\left(x_2\right)\overset{x_{1,2} \to x_0}{\simeq} \hat{C}_{\phi\phi}^{\mathbb{1}}\left(x_1, x_2; x_0\right)\mathbb{1} + \hat{C}_{\phi\phi}^{\phi}\left(x_1, x_2; x_0\right)\phi(x_0)+ \dots,
\end{equation}
where $\hat{C}_{\phi\phi}^{\phi}$ is proportional to the (generically dimensionful) cubic coupling $\lambda_3$. In the presence of a $\mathbb{Z}_2$ symmetry, on the other hand, the simplest interaction is quartic, that in the OPE language contributes to the coefficient $\hat{C}_{\phi\phi}^{\phi^2}$:
\begin{equation}
\mathcal{\phi}\left(x_1\right)\mathcal{\phi}\left(x_2\right)\overset{x_{1,2} \to x_0}{\simeq} \hat{C}_{\phi\phi}^{\mathbb{1}}\left(x_1, x_2; x_0\right)\mathbb{1} + \hat{C}_{\phi\phi}^{\phi^2}\left(x_1, x_2; x_0\right)\phi^2(x_0)+ \dots,
\end{equation}
where $\phi^2(x)$ is a composite operator and $\hat{C}_{\phi\phi}^{\phi^2}$ is proportional to the quartic coupling $\lambda$.

\subsection{Cubic Interaction}
To compute $\hat{C}_{\phi\phi}^{\phi}$, we consider a $\phi^3$ theory and examine the three-point correlation function: 
\begin{equation} 
\expval{\phi\left(x_1\right)\phi\left(x_2\right)\phi\left(x_3\right)}_g. 
\end{equation} 
By taking a hierarchical limit in which we first take $x_{1,2} \to x_0$ and then $x_{3} \to x_0$, we can decompose the three-point correlation function into a product of OPE coefficients. This approach will allow us to extract $\hat{C}_{\phi\phi}^{\phi}$ and to also perform a useful consistency check, by independently reproducing the coefficient $\hat{C}_{\phi\phi}^{\mathbb{1}}$ that we discussed in the previous section.  

In more detail, we first take $x_1$ and $x_2$ to approach the midpoint $x_0= (x_1 + x_2)/2$ while holding $x_3$ fixed. At leading order, this corresponds to isolating the contribution from the~$\hat{C}_{\phi\phi}^{\phi}$ OPE coefficient in the tree-point function: 
\begin{equation} 
\expval{\phi\left(x_1\right)\phi\left(x_2\right)\phi\left(x_3\right)}_g \overset{x_{1,2} \to x_0}{\simeq} \hat{C}_{\phi\phi}^{\phi}(\Delta x, x_0) \expval{\phi\left(x_0\right)\phi\left(x_3\right)}_g. 
\end{equation}
where we expressed $\hat{C}_{\phi\phi}^{\phi}$ in terms of the separation $\Delta x = x_2 - x_1$ and the midpoint $x_0$. 
Next, we take the limit $x_{3} \to x_0$ which (again at leading order) selects the OPE coefficient $\hat{C}_{\phi\phi}^{\mathbb{1}}$, conveniently expressed as a function of $y = x_3 - x_0$ and $x_0$:
\begin{equation} 
\lim_{x_{3} \to x_0}\lim_{x_{1,2} \to x_0}\expval{\phi\left(x_1\right)\phi\left(x_2\right)\phi\left(x_3\right)}_g \simeq \hat{C}_{\phi\phi}^{\phi}(\Delta x, x_0) 
\hat{C}_{\phi\phi}^{\mathbb{1}}\left(y, x_0\right). 
\end{equation}
The second term in this product will be seen to match Eq.~\eqref{eq:2ptHadamard_OPE}. We are now interested in determining the gravitational corrections in the OPE coefficient $\hat{C}_{\phi\phi}^{\phi}$. To isolate this contribution, it is convenient to expand this OPE coefficient perturbatively in orders of the metric perturbation, as we previously did for the Green's function: 
\begin{equation}
\label{eq:OPE Grav Expansion}
\hat{C}_{\phi\phi}^{\phi}\left(\Delta x, x_0\right) = \hat{C}_{\phi\phi}^{\phi}\0\left(\Delta x\right) + \hat{C}_{\phi\phi}^{\phi}\1\left(\Delta x, x_0\right)+\dots.
\end{equation}
Note that the flat space term only depends on $\Delta x$ since it enjoys full Lorentz invariance, whereas the gravitational correction breaks translation invariance and depends also on $x_0$.

With this framework established, we proceed to compute the relevant terms by first writing down the interaction action: 
\begin{equation}
S_{\text{int}} = \frac{1}{2} \int \bar{h}_{\mu\nu} \partial^\mu \phi \partial^\nu \phi \dd[\n]{x} + \frac{1}{3!}\int \lambda_3 \left(\phi^3 - \frac{1}{\n - 2}\bar{h}\phi^3\right)\dd[\n]{x}, 
\end{equation} 
where the first term arises from the kinetic part of the action and $\bar{h}_{\mu\nu}$ is defined as
\begin{equation}
\label{eq:hbar}
\bar{h}_{\mu\nu} = h_{\mu\nu} - \frac{1}{2}h \eta_{\mu\nu}.
\end{equation}
Note that its trace is $\bar{h} = \left(1 - d/2\right)h$. For simplicity we discuss the case of minimal coupling, but the analysis is easily modified for nonminimal couplings. The first term in the interaction action is given by the vertex
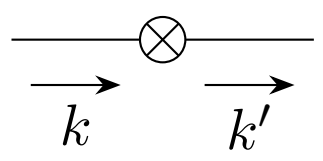
\begin{equation}
\label{eq:hinteractionvertex}
    \begin{gathered}
    \vspace{-0.5cm}
        \feynmandiagram[horizontal=a to c, layered layout]
        {a -- [plain, momentum'={$k$}] b[crossed dot, minimum size=3mm] -- [plain, momentum'={$k'$}] c};
    \end{gathered} = \frac{i}{2}\left(k^\mu k'^\nu + k'^\mu k^\nu\right)\bar{h}_{\mu\nu}\left(k' - k\right),
\end{equation}
Meanwhile, the remaining two terms in the interaction action give rise to
\begin{subequations}
\begin{align}
\begin{gathered}
\begin{tikzpicture}[scale=0.75]
\begin{feynman}
\vertex (a) at (-1, 1);
\vertex (b) at (-1, -1);
\vertex (c) at (0, 0);
\vertex (d) at (1.5, 0);
\diagram*{(a) -- [plain, momentum=$k_1$] (c) -- [plain, momentum'=$k_3$] (d);
(b) -- [plain, momentum'=$k_2$] (c);};
\end{feynman}
\end{tikzpicture}
\end{gathered} &= i\lambda_3 \left(2\pi\right)^\n \delta^\n\left(k_3 - k_1 - k_2\right),\\
\begin{gathered}
\begin{tikzpicture}[scale=0.75]
\begin{feynman}
\vertex (a) at (-1, 1);
\vertex (b) at (-1, -1);
\vertex (c) at (0, 0);
\vertex (d) at (1.5, 0);
\node [crossed dot, minimum size = 3mm, draw, thick] (e) at (0, 0);
\diagram*{(a) -- [plain, momentum=$k_1$] (e) -- [plain, momentum'=$k_3$] (d);
(b) -- [plain, momentum'=$k_2$] (c);};
\end{feynman}
\end{tikzpicture}
\end{gathered} &= -\frac{i\lambda_3}{\n - 2}{\bar{h}}\left(k_3 - k_1 - k_2\right).
\end{align}
\end{subequations}

Let us organize the diagrams contributing at first-order in $h$ by their order in the gravitational correction to the coefficient $\hat{C}_{\phi\phi}^{\phi}$. At lowest order, we have at most one gravitational insertion on the leg corresponding to the propagator from $x_0$ to $x_3$:
\begin{equation} 
\label{eq:zeroth_diagram}
\hat{C}_{\phi\phi}^{\phi}\0\left(\Delta x\right) \hat{C}_{\phi\phi}^{\mathbb{1}}\left(y, x_0\right) \simeq 
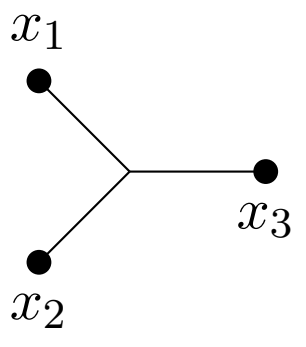
\begin{gathered}
\begin{tikzpicture}[scale=0.6]
\begin{feynman}
\vertex (a) at (-1, 1);
\vertex (b) at (-1, -1);
\vertex (c) at (0, 0);
\vertex (d) at (1.5, 0);
\node [dot, label=above:{$x_1$}]  at (-1, 1);
\node [dot, label=below:{$x_2$}] at (-1, -1);
\node [dot, label=below:{$x_3$}] at (1.5, 0);
\diagram*{(a) -- [plain] (c) -- [plain] (d);
(b) -- [plain] (c);};
\end{feynman}
\end{tikzpicture}
\end{gathered} + 
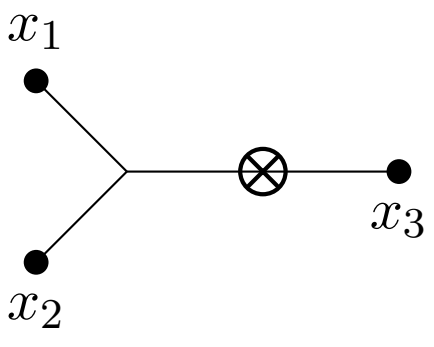
\begin{gathered}
\begin{tikzpicture}[scale=0.6]
\begin{feynman}
\vertex (a) at (-1, 1);
\vertex (c) at (0, 0);
\vertex (b) at (-1, -1);
\vertex (e) at (1.5, 0);
\vertex (d) at (3, 0);
\node [dot, label=above:{$x_1$}] at (-1, 1);
\node [crossed dot, minimum size=3mm, thick] at (1.5, 0);
\node [dot, label=below:{$x_3$}] at (3, 0);
\node [dot, label=below:{$x_2$}] at (-1, -1);
\diagram*{(a) -- [plain] (c) -- (e) -- [plain] (d);
(b) -- [plain] (c);};
\end{feynman}
\end{tikzpicture}
\end{gathered},
\end{equation}
where the hierarchical limits previously discussed are intended. The remaining diagrams, which contain gravitational insertions on the $x_1$ and $x_2$ legs, or on the cubic vertex, contribute to the leading gravitational corrections in $\hat{C}_{\phi\phi}^{\phi}$. At this order, the propagator from $x_0$ to $x_3$ is the flat space one. Diagrammatically we have
\begin{equation}
\label{eq:2ndsetofdiagrams}
\begin{aligned}
\hat{C}_{\phi\phi}^{\phi}\1\left(\Delta x, x_0\right)\hat{C}_{\phi\phi}^{\mathbb{1}}\0\left(y\right) &\simeq 
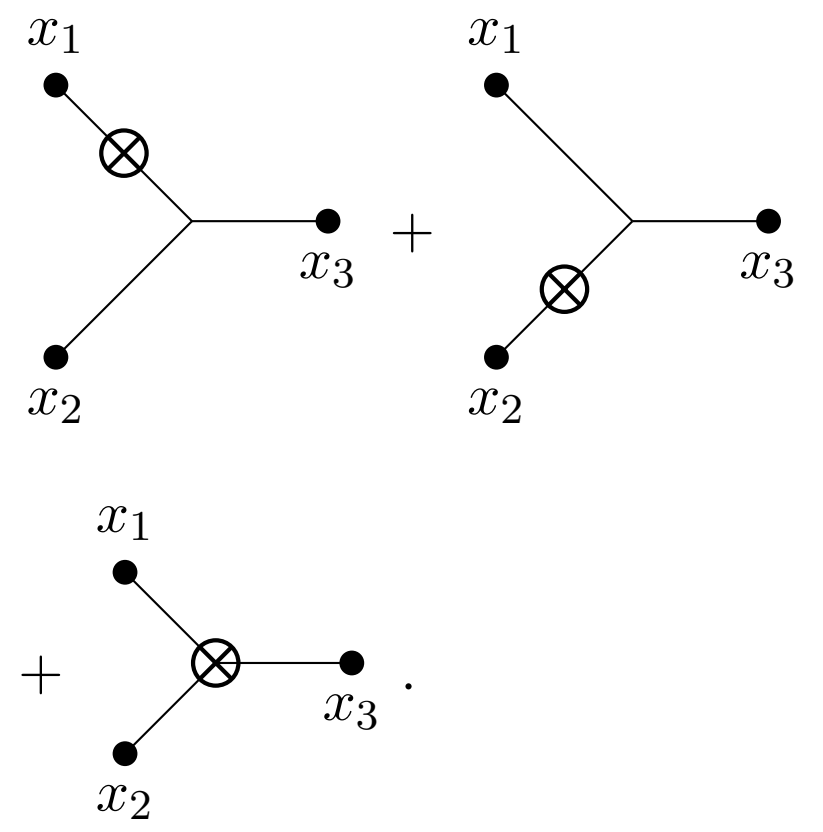
\begin{gathered}
\begin{tikzpicture}[scale=0.6]
\begin{feynman}
\vertex (a) at (-1.5, 1.5);
\vertex (e) at (-0.75, 0.75);
\vertex (b) at (-1.5, -1.5);
\vertex (c) at (0, 0); 
\vertex (d) at (1.5, 0);
\node [dot, label=above:{$x_1$}] at (-1.5, 1.5);
\node [crossed dot, minimum size=3mm, thick] at (-0.75, 0.75);
\node [dot, label=below:{$x_3$}] at (1.5, 0);
\node [dot, label=below:{$x_2$}] at (-1.5, -1.5);
\diagram*{(a) -- (e) -- [plain] (c) -- [plain] (d);
(b) -- [plain] (c);};
\end{feynman}
\end{tikzpicture}
\end{gathered} +
\begin{gathered}
\begin{tikzpicture}[scale=0.6]
\begin{feynman}
\vertex (a) at (-1.5, 1.5);
\vertex (b) at (-1.5, -1.5);
\vertex (e) at (-0.75, -0.75);
\vertex (c) at (0, 0);
\vertex (d) at (1.5, 0);
\node [dot, label=above:{$x_1$}] at (-1.5, 1.5);
\node [dot, label=below:{$x_3$}] at (1.5, 0);
\node [dot, label=below:{$x_2$}] at (-1.5, -1.5);
\node [crossed dot, minimum size=3mm, thick] at (-0.75, -0.75);
\diagram*{(a) -- [plain] (c) -- [plain] (d);
(b) -- (e) -- [plain] (c);};
\end{feynman}
\end{tikzpicture}
\end{gathered} \\
&\hspace{0.5cm} + 
\begin{gathered}
\begin{tikzpicture}[scale=0.6]
\begin{feynman}
\vertex (a) at (-1, 1);
\vertex (b) at (-1, -1);
\vertex (c) at (0, 0);
\vertex (d) at (1.5, 0);
\node [dot, label=above:{$x_1$}] at (-1, 1);
\node [crossed dot, minimum size=3mm, thick] at (0, 0);
\node [dot, label=below:{$x_3$}] at (1.5, 0);
\node [dot, label=below:{$x_2$}] at (-1, -1);
\diagram*{(a) -- [plain] (c) -- [plain] (d);
(b) -- [plain] (c);};
\end{feynman}
\end{tikzpicture}
\end{gathered}.
\end{aligned}
\end{equation}

Having organized the diagrams, we proceed to compute them accordingly. The computation is most conveniently performed by changing variables from $(x_1, x_2, x_3)$ to $(\Delta x, x_0, y)$:
\begin{equation}\label{eq:x3}
\begin{pmatrix}
\Delta x \\
x_0 \\ 
y \\
\end{pmatrix} 
=
\begin{pmatrix}
1 & -1 & 0  \\
1/2 & 1/2  & 0  \\
-1/2 & -1/2  & 1  \\
\end{pmatrix} 
\begin{pmatrix}
x_1 \\
x_2 \\ 
x_3 \\
\end{pmatrix} \, .
\end{equation}
Correspondingly, we perform a change of variable on the momenta in such a way that the Fourier transform factors are factorized\footnote{In our convention the momenta $k_1,k_2$ are incoming, while $k_3$ is outgoing.}
\begin{equation}\label{eq:k3}
\begin{pmatrix}
p \\
q \\ 
k \\
\end{pmatrix} 
=
\begin{pmatrix}
1/2 & -1/2 & 0  \\
1 & 1  & 1  \\
0 & 0  & 1  \\
\end{pmatrix} 
\begin{pmatrix}
- k_1 \\
- k_2 \\ 
k_3 \\
\end{pmatrix} \, .
\end{equation}
In this way the Fourier transform in the new variables is in a factorized form: 
\begin{equation}
\int_{k_1, k_2, k_3} e^{-ik_1 x_1 - i k_2 x_2 + i k_3 x_3} = \int_{p,q,k} e^{i p \Delta x} e^{i q x_0} e^{i k y}
\end{equation}
The Jacobian factor of this transformation is $1$. These momenta have a nice physical interpretation: $q$ is the momentum carried by the graviton, while $p$ and $k$ are relatively hard momenta. These variables make manifest the hierarchical OPE expansion and, importantly, simplify the computation by factorizing the integrals. There are two expansions to do under the integral. The first expansion isolates terms of second order in $q/p$ when the gravitational vertex is on the $x_1$ or $x_2$ leg, and in $q/k$ when the gravitational vertex is on the $x_3$ leg. The second approximation is to take $p \gg k$, corresponding to the fact that $x_3$ approaches $x_0$ more slowly than $x_1$ and $x_2$. Therefore we expand to the lowest order in $k/p$. By doing this we are implicitly assuming that the position-space OPE can be obtained as the Fourier transform of the momentum-space OPE. This holds in perturbation theory as long as the operators with small momentum are elementary fields, or do not have a large scaling dimension~\cite{Collins:1984xc,Dymarsky:2014zja}. We will restrict our analysis to this case in what follows.

For the diagrams in Eq.~\eqref{eq:zeroth_diagram}, we have:
\begin{equation}
\label{zeroth_interation}
\begin{aligned}
\hat{C}_{\phi\phi}^{\phi}\0\left(\Delta x\right) C_{\phi\phi}^{\mathbb{1}}\left(y, x_0\right) = &\lambda_3 \int_{k_1, k_2, k_3} \Bigg(\frac{\left(-i\right)^2 \left(2\pi\right)^\n\delta^\n\left(q\right)}{\left(k_1^2 - i\epsilon\right)\left(k_2^2 - i\epsilon\right)\left(k_3^2 - i\epsilon\right)} \\
&+ \frac{\left(1/2\right)\left(-i\right)^2\left(k_3^\mu \left(k_3 - q\right)^\nu + k_3^\nu \left(k_3 - q\right)^\mu\right)}{\left(k_1^2 - i\epsilon\right)\left(k_2^2 - i\epsilon\right)\left(\left(k_3 - q\right)^2 - i\epsilon\right)\left(k_3^2 - i\epsilon\right)} {\bar{h}}_{\mu\nu}\left(q\right)\Bigg) \times \\
& e^{-ik_1 x_1} e^{-ik_2 x_2} e^{ik_3 x_3}.
\end{aligned}
\end{equation}
In writing this, the integral of the cubic vertex delta function has already been evaluated in the second term.
After changing variables and expanding, the result factorizes into two integrals:
\begin{equation}
\hat{C}_{\phi\phi}^{\phi}\0\left(\Delta x\right) = \lambda_3\int_{p} \frac{-i}{\left(p^2 - i\epsilon\right)^2} e^{ip\Delta x}
\end{equation}
and 
\begin{equation}
\begin{aligned}
\hat{C}_{\phi\phi}^{\mathbb{1}}\left(y, x_0\right) = \int_{q, k} &\Bigg(\frac{-i\left(2\pi\right)^\n\delta^\n\left(q\right)}{k^2 - i\epsilon} - \frac{-i{\bar{h}}_{\mu\nu}\left(q\right)}{\left(k^2 - i\epsilon\right)^3}\times \\
&\hspace{0.2cm}\left(q^2 k^\mu k^\nu - 2\left(k \cdot q\right)k^{(\mu} q^{\nu)} + \frac{4\left(k\cdot q\right)^2 k^\mu k^\nu}{k^2 - i\epsilon}\right)e^{iqx_0} e^{iky}.
\end{aligned}
\end{equation}
The first integral gives: 
\begin{equation}\label{eq:cubic0}
\hat{C}_{\phi\phi}^{\phi}\0\left(\Delta x\right) = \lambda_3 \, I_2\left(\Delta x\right),
\end{equation}
with $I_2\left(\Delta x\right)$ defined in Eq.~\eqref{eq:GenericI}. 
On the other hand, performing the second integral we find that $C_{\phi\phi}^{\mathbb{1}}\left(y, x_0\right)$ reproduces Eq.~\eqref{eq:2ptHadamard_OPE}, with the normalization $C_d = \Gamma\left(d/2 - 1\right)/\left(4\pi^{d/2}\right)$. The equality is not manifest at the integrand level, and this therefore constitutes a non-trivial consistency check of the formalism.

Now we turn our attention to the second set of diagrams in Eq.~\eqref{eq:2ndsetofdiagrams}, which once again factorize. First,
\begin{equation}
\hat{C}_{\phi\phi}^{\mathbb{1}}\0\left(y\right) = \int_{k} \frac{-i}{k^2 - i\epsilon} e^{iky}.
\end{equation}
The second factor is as follows, with the second order in $\left(q/p\right)$ isolated:
\begin{equation}
\begin{aligned}
\hat{C}_{\phi\phi}^{\phi}\1\left(\Delta x, x_0\right) &= \lambda_3 \int_{p, q} \frac{-i{\bar{h}}\left(q\right)}{\left(p^2 - i\epsilon\right)^3} \Bigg(-\frac{1}{2}q^\mu q^\nu - \frac{3 q^2 p^\mu p^\nu}{2\left(p^2 - i\epsilon\right)} \\
&\hspace{0.5cm} + \frac{4\left(p \cdot q\right)^2 p^\mu p^\nu}{\left(p^2 - i\epsilon\right)^2} + \frac{\eta^{\mu\nu} q^2}{2\left(\n - 2\right)} - \frac{\eta^{\mu\nu}\left(p\cdot q\right)^2}{\left(\n - 2\right)\left(p^2 - i\epsilon\right)}\Bigg)e^{ip\Delta x} e^{iqx_0}.
\end{aligned}
\end{equation}
As before, this can be expressed in terms of derivatives of $I_\alpha$, 
\begin{equation}\label{eq:cubic1}
\begin{aligned}
\hat{C}_{\phi\phi}^{\phi}\1\left(\Delta x, x_0\right) &= \frac{\lambda_3}{2}I_3\left(\Delta x\right)\partial^\mu \partial^\nu \bar{h}_{\mu\nu}\left(x_0\right) - \lambda_3\frac{3}{2}\partial^\mu \partial^\nu I_4\left(\Delta x\right)\Box\bar{h}_{\mu\nu}\left(x_0\right)\\
&\hspace{0.5cm} - 4 \lambda_3 \, \partial^\rho \partial^\sigma \partial^\mu \partial^\nu I_5\left(\Delta x\right)\partial_\rho \partial_\sigma \bar{h}_{\mu\nu}\left(x_0\right)\\
&\hspace{0.5cm} -\frac{\lambda_3}{2\left(\n - 2\right)}I_3\left(\Delta x\right)\Box\bar{h}\left(x_0\right) - \frac{\lambda_3}{\n - 2}\partial^\rho \partial^\sigma I_4\left(\Delta x\right)\partial_\rho\partial_\sigma \bar{h}\left(x_0\right).
\end{aligned}
\end{equation}
Combining the zeroth-order and first-order results in Eq.~\eqref{eq:cubic0} and Eq.~\eqref{eq:cubic1}  yields
\begin{equation}
\begin{aligned}
\hat{C}_{\phi\phi}^{\phi}\left(\Delta x, x_0\right) = \lambda_3\frac{\Gamma\left(\n/2 - 2\right)}{16\pi^{\n/2}\left(\Delta x^2 + i\epsilon\right)^{\n/2 - 3}}&\Bigg(\frac{1}{\Delta x^2 + i\epsilon}\left(1 + \frac{\Delta x^\mu \Delta x^\nu}{12} R_{\mu\nu}\left(x_0\right)\right) \\
&+ \frac{1}{12\left(\n - 6\right)}R\left(x_0\right) \\
&- \frac{\left(\n - 4\right)\Delta x^\mu \Delta x^\nu \Delta x^\rho \Delta x^\sigma}{48\left(\Delta x^2 + i\epsilon\right)^2}\partial_\rho\partial_\sigma h_{\mu\nu}\left(x_0\right)\Bigg).
\end{aligned}
\end{equation}
We can express this result in terms of geometric invariants by using the expansions in Eq.~\eqref{eq:SigmaExp} and Eq.~\eqref{eq:vvexpansion}, obtaining the leading gravitational correction to the $\hat{C}_{\phi\phi}^{\phi}$ OPE coefficient:
\begin{equation}
\label{eq:FinalCubicResult}
\hat{C}_{\phi\phi}^{\phi}\left(\Delta x, x_0\right) = \lambda_3\frac{\Gamma\left(\n/2 - 2\right)}{16\pi^{\n/2}\left(s^2 + i\epsilon\right)^{\n/2 - 3}}\left(\frac{\mathcal{D}^{1/2}_{12}}{s^2 + i\epsilon} + \frac{1}{12\left(\n - 6\right)}R\left(x_0\right)\right),
\end{equation}
where $\mathcal{D}_{12} \equiv \mathcal{D}\left(x_1, x_2\right)$.

In the four-dimensional limit, setting $\n = 4 - 2\epsilon$ and taking $\epsilon$ to $0$, we find
\begin{equation}
\lim_{\n\rightarrow 4}\hat{C}_{\phi\phi}^{\phi}\left(\Delta x, x_0\right) = -\frac{\lambda_3}{16\pi^2}\left(\mathcal{D}^{1/2}_{12} - \frac{s^2}{24}R\left(x_0\right)\right)\log\left(s^2 + i\epsilon\right),
\end{equation}
up to analytic terms. The non-analytic part includes a $\log(s^2)$ singularity multiplying the van Vleck determinant, $\mathcal{D}$, at leading order. Additionally, there is a gravitational correction proportional to $s^2 \log(s^2) R$, which introduces a new nonanalytic structure. 

\subsection{Quartic Interaction}
We now extend the previous analysis to a weakly coupled theory with a quartic coupling, $\lambda$. We are interested in computing gravitational corrections to the OPE coefficient $\hat{C}_{\phi\phi}^{\phi^2}\left(x_1, x_2; x_0\right)$. For simplicity, we will take the theory to have $\mathbb{Z}_2$ symmetry, such that the OPE is simply
\begin{equation}
\mathcal{\phi}\left(x_1\right)\mathcal{\phi}\left(x_2\right)\overset{x_{1,2} \to x_0}{\simeq} \hat{C}_{\phi\phi}^{\mathbb{1}}\left(x_1, x_2; x_0\right)\mathbb{1} + \hat{C}_{\phi\phi}^{\phi^2}\left(x_1, x_2; x_0\right)\phi^2(x_0)+ \dots,
\end{equation}
In analogy to the previous case, we consider consider a four-point correlation function and take a hierarchical limit in which we first send $x_{1,2} \to x_0$, and only after we take $x_{3,4} \to x_0$. Using the OPE for $\phi(x_1)$ and $\phi(x_2)$, working at leading order and discarding any disconnected correlators, we find
\begin{equation}
\expval{\phi\left(x_1\right)\phi\left(x_2\right)\phi\left(x_3\right)\phi\left(x_4\right)} \overset{x_{1,2} \to x_0}{\simeq} \hat{C}_{\phi\phi}^{\phi^2}\left(\Delta x, x_0\right)\expval{\phi^2\left(x_0\right)\phi\left(x_3\right)\phi\left(x_4\right)}.
\end{equation}
We next take send $x_3$ and $x_4$ to $x_0$. Again working at leading order in the coupling and dropping disconneted terms, we obtain
\begin{equation}
\lim_{x_{3,4} \to x_0}\lim_{x_{1,2} \to x_0}\expval{\phi\left(x_1\right)\phi\left(x_2\right)\phi\left(x_3\right)\phi\left(x_4\right)} \simeq 2 \hat{C}_{\phi\phi}^{\phi^2}\left(\Delta x, x_0\right)\hat{C}_{\phi\phi}^{\mathbb{1}}\left(y, x_0\right)\hat{C}_{\phi\phi}^{\mathbb{1}}\left(z, x_0\right),
\end{equation}
where $y = x_3 - x_0$ and $z = x_4 - x_0$, and the factor of two is a combinatorial factor. Finally, we expand the yet undetermined OPE coefficient to first order in gravitational corrections:
\begin{equation}
\expval{\phi\left(x_1\right)\phi\left(x_2\right)\phi\left(x_3\right)\phi\left(x_4\right)} = 2 \left(\hat{C}_{\phi\phi}^{\phi^2}\0\left(\Delta x\right) + \hat{C}_{\phi\phi}^{\phi^2}\1\left(\Delta x, x_0\right)\right)\hat{C}_{\phi\phi}^{\mathbb{1}}\left(y, x_0\right)\hat{C}_{\phi\phi}^{\mathbb{1}}\left(z, x_0\right).
\end{equation}

For concreteness, consider a minimally coupled scalar with $\phi^4$ interaction:
\begin{equation}
S_{\text{int}} = \frac{1}{2}\int \bar{h}_{\mu\nu} \partial^\mu \phi \partial^\nu \phi \dd[\n]{x} - \frac{1}{4!}\int \lambda\left(\phi^4 - \frac{1}{\n - 2}\bar{h}\phi^4\right)\dd[\n]{x},
\end{equation}
with $\bar{h}$ defined in Eq.~\eqref{eq:hbar}. The analysis can be easily extended to more general interactions and nonminimal couplings. The new momentum space vertices, with $k_1$ and $k_2$ incoming and $k_3$ and $k_4$ outgoing, are
\begin{subequations}
\begin{align}
\begin{gathered}
\begin{tikzpicture}[scale=0.6]
\begin{feynman}
\vertex (a) at (-1, 1);
\vertex (b) at (-1, -1);
\vertex (d) at (1, 1);
\vertex (e) at (1, -1);
\vertex (c) at (0, 0);
\diagram*{(a) -- [plain] (c) -- [plain] (d);
(b) -- [plain] (c);
(c) -- [plain] (e);};
\end{feynman}
\end{tikzpicture}
\end{gathered} &\:= -i\lambda \left(2\pi\right)^\n \delta^\n\left(k_4 + k_3 - k_1 - k_2\right),\\
\begin{gathered}
\begin{tikzpicture}[scale=0.6]
\begin{feynman}
\vertex (a) at (-1, 1);
\vertex (b) at (-1, -1);
\vertex (d) at (1, 1);
\vertex (e) at (1, -1);
\vertex (c) at (0, 0);
\node [crossed dot, minimum size = 3mm, draw, thick] at (0, 0);
\diagram*{(a) -- [plain] (c) -- [plain] (d);
(b) -- [plain] (c);
(c) -- [plain] (e);};
\end{feynman}
\end{tikzpicture}
\end{gathered} &\:= \frac{i\lambda}{\n - 2}{\bar{h}}\left(k_4 + k_3 - k_1 - k_2\right).
\end{align}
\end{subequations}

We can once again organize the contributing diagrams by their order in gravitational corrections to $\hat{C}_{\phi\phi}^{\phi^2}\left(\Delta x, x_0\right)$ and $\hat{C}_{\phi\phi}^{\mathbb{1}}$. In practice, this means separating the diagrams into those that have a vertex on the $x_1$ or $x_2$ leg and those that do not. To begin, the zeroth-order contribution to $\hat{C}_{\phi\phi}^{\phi^2}\left(\Delta x\right)$ can be extracted from
\begin{equation} 
\begin{aligned}
2\hat{C}_{\phi\phi}^{\phi^2}\0\left(\Delta x\right)\hat{C}_{\phi\phi}^{\mathbb{1}}\0\left(y\right) \hat{C}_{\phi\phi}^{\mathbb{1}}\0\left(z\right) = 
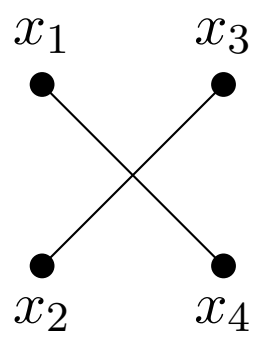
\begin{gathered}
\begin{tikzpicture}[scale=0.6]
\begin{feynman}
\vertex (a) at (-1, 1);
\vertex (b) at (-1, -1);
\vertex (c) at (0, 0);
\vertex (d) at (1, 1);
\vertex (e) at (1, -1);
\node [dot, label=above:{$x_1$}]  at (-1, 1);
\node [dot, label=above:{$x_3$}] at (1, 1);
\node [dot, label=below:{$x_2$}] at (-1, -1);
\node [dot, label=below:{$x_4$}] at (1, -1);
\diagram*{(a) -- [plain] (c) -- [plain] (d);
(b) -- [plain] (c);
(c) -- [plain] (e);};
\end{feynman}
\end{tikzpicture}
\end{gathered}.
\end{aligned}
\end{equation}
The integral expression takes the form
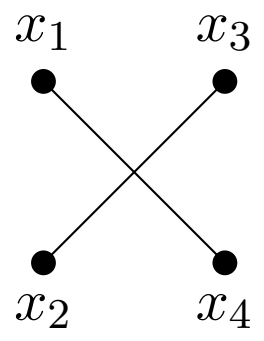
\begin{equation}
\begin{gathered}
\begin{tikzpicture}[scale=0.6]
\begin{feynman}
\vertex (a) at (-1, 1);
\vertex (b) at (-1, -1);
\vertex (c) at (0, 0);
\vertex (d) at (1, 1);
\vertex (e) at (1, -1);
\node [dot, label=above:{$x_1$}]  at (-1, 1);
\node [dot, label=above:{$x_3$}] at (1, 1);
\node [dot, label=below:{$x_2$}] at (-1, -1);
\node [dot, label=below:{$x_4$}] at (1, -1);
\diagram*{(a) -- [plain] (c) -- [plain] (d);
(b) -- [plain] (c);
(c) -- [plain] (e);};
\end{feynman}
\end{tikzpicture}
\end{gathered} \: \: = \int_{k_1, k_2, k_3, k_4} \frac{(-i\lambda)\left(2\pi\right)^4\delta^4\left(q\right)e^{-ik_1 x_1} e^{-ik_2 x_2} e^{ik_3 x_3} e^{ik_4 x_4}}{\left(k_1^2 - i\epsilon\right)\left(k_2^2 - i\epsilon\right)\left(k_3^2 - i\epsilon\right)\left(k_4^2 - i\epsilon\right)},
\end{equation}
where $q$ will be defined shortly. As before, we can switch the position variables to the natural variable for the OPE expansion under consideration. $\Delta x$, $x_0$, and $y$ are defined as before in~Eq.~\eqref{eq:x3}, while $z = x_4 - x_0$. Similarly, we choose momenta in such a way that the Fourier transform keeps a factorized form. The new momenta $p$ and $k$ are defined as in~Eq.~\eqref{eq:k3}, while \begin{equation}
q = k_1 + k_2 - k_3 - k_4 \qquad {\rm and} \qquad \ell = k_4.
\end{equation}
Performing this transformation, integrating in $q$ and expanding to lowest order in both $\ell/p$ and $k/p$, we find that the integrals factorize. As expected, the $k$ and $\ell$ integrals reproduce  $\hat{C}_{\phi\phi}^{\mathbb{1}}\0$ for $y$ and $z$ respectively, while the remaining $p$ integral gives
\begin{equation}
\hat{C}_{\phi\phi}^{\phi^2}\0\left(\Delta x\right) = -\frac{\lambda}{2}I_2\left(\Delta x\right).
\end{equation}
Next there are two diagrams with gravitational background insertions on the $x_3$ or $x_4$ leg. As before, they contribute gravitational corrections to $\hat{C}_{\phi\phi}^{\mathbb{1}}$, matching our previous computation of the Hadamard tail.

Finally we can turn our attention to the diagrams with gravitational insertions on the $x_1$ and $x_2$ legs, and in the quartic vertex. The calculation proceeds analogously to the previous case, and the propagators associated to $k$ and $\ell$ factorize. The remaining integral for $\hat{C}_{\phi\phi}^{\phi^2}\1$ is the same as the one for $\hat{C}_{\phi\phi}^{\phi}\1$ in the cubic case, up to a combinatorial factor of $1/2$ as occurred in the zeroth-order case. In other words, for this theory, $\hat{C}_{\phi\phi}^{\phi^2}$ is $1/2$ times the result in Eq.~\eqref{eq:FinalCubicResult}, with the obvious substitution $\lambda_3\to -\lambda$. Therefore we arrive at
\begin{equation}\label{eq:C4minimal}
\hat{C}_{\phi\phi}^{\phi^2}\left(\Delta x, x_0\right) = -\lambda \,\frac{\Gamma\left(\n/2 - 2\right)}{32\pi^{\n/2}\left(s^2 + i\epsilon\right)^{\n/2 - 3}}\left(\frac{\mathcal{D}^{1/2}_{12}}{s^2 + i\epsilon} + \frac{1}{12\left(\n - 6\right)}R\left(x_0\right)\right).
\end{equation}
In four dimensions, setting $\n = 4 - 2\epsilon$ and taking $\epsilon$ to $0$, we find
\begin{equation}\label{eq:phi4d4}
\lim_{\n\rightarrow 4}\hat{C}_{\phi\phi}^{\phi^2}\left(\Delta x, x_0\right) = \frac{\lambda}{32\pi^2}\left(\mathcal{D}^{1/2}_{12} - \frac{s^2}{24}R\left(x_0\right)\right)\log\left(s^2 + i\epsilon\right),
\end{equation}
In the flat space limit this result agrees with the position space OPE coefficient computed in~\cite[section 10.1.2]{Collins:1984xc}.

\subsection{The Wilson-Fisher model}

It is interesting to note that $\gamma_{\phi^2}=\lambda/16\pi^2$ is the one-loop anomalous dimension of the composite operator $\phi^2(x)$ in the massless $\lambda \phi^4$ theory, so that at leading order in $\lambda$ we can rewrite our result of Eq.~\eqref{eq:phi4d4} as
\begin{equation}
\lim_{\n\rightarrow 4}\hat{C}_{\phi\phi}^{\phi^2}\left(\Delta x, x_0\right) = \left(\mathcal{D}^{1/2}_{12} - \frac{s^2}{24}R\left(x_0\right)\right)s^{\gamma_{\phi^2}}.
\end{equation}
This observation is suggestive, but to make it more precise we can consider the Wilson-Fisher fixed point in the $\epsilon$ expansion~\cite{Wilson:1971dc,Wilson:1972cf}. We work in $d= 4- 2\epsilon$ dimensions, where $\epsilon$ is a small but finite perturbative parameter, and consider the theory with action
\begin{equation}
S =  \int \left(-\frac{1}{2} g^{\mu\nu}\partial_\mu \phi\partial_\nu \phi -\frac{\xi}{2}R \,\phi^2 - \dfrac{\lambda}{4!}\phi^4 \right)\sqrt{-g} \dd[\n]{x}.
\end{equation}
In flat space, this theory is a CFT if the quartic is chosen to satisfy $3\lambda/32\pi^2 = \epsilon$. We will assume this choice from now on in this section. The scaling dimensions of the operators $\phi$ and $\phi^2$ at the fixed point are~\cite{Wilson:1972cf}
\begin{subequations}
\begin{align}
\label{eq:WF Deltaphi}
\Delta_\phi &= 1- \epsilon \, , \\
\Delta_{\phi^2} &= 2 - 2\epsilon +\gamma_{\phi^2}\, .
\label{eq:WF Deltaphisq}
\end{align}
\end{subequations}
Conformal invariance dictates the scaling behavior of the OPE coefficient $C_{\phi\phi}^{\phi^2}$ to be 
\begin{equation}
C_{\phi\phi}^{\phi^2}(x) \propto x^{\Delta_{\phi^2} - 2\Delta_\phi} = x^{\gamma_{\phi^2}},
\end{equation}
which is consistent with our previous result in massless $\lambda \phi^4$.

Moving away from flat space, and working at leading order in $\lambda$, we notice that this theory is also classically Weyl invariant if we choose conformal coupling, 
\begin{equation}
\xi = \frac{d - 2}{4\left(d - 1\right)}.
\end{equation}
The two point function is given in this limit by \cite{Chavda:2025ijz}
\begin{equation}
\label{eq:nonmin_GF}
\expval{\phi\left(x_1\right)\phi\left(x_2\right)} = \frac{\Gamma\left(d/2 - 1\right)}{4\pi^{d/2}\left(s^2 + i\varepsilon\right)^{d/2 - 1}}\left(\mathcal{D}_{12}^{1/2} - \frac{s^2 R\left(x_0\right)}{24\left(d - 1\right)} \right).
\end{equation}
For the gravitational correction to the OPE coefficient $\hat{C}_{\phi\phi}^{\phi^2}$ we can start from the result Eq.~\eqref{eq:C4minimal}, obtained for minimal coupling, and add the contributions from the non-minimal scalar curvature coupling. Keeping generic $d$ and $\xi$ for the moment we have
\begin{equation}
2 \,\delta \hat{C}_{\phi\phi}^{\phi^2}(\Delta x) = 2 (-i\lambda) \int_p \Big(-i \xi R(x_0) \Big) \frac{(-i)^3}{(p^2-i\epsilon)^3} e^{i p \Delta x},
\end{equation}
where the factor of $2$ on the right comes from a sum over insertions on the two legs. Carrying out the computation we find that for nonminimal coupling
\begin{equation}
\label{eq:C4nonminimal}
\hat{C}_{\phi\phi}^{\phi^2}\left(\Delta x, x_0\right) = -\lambda \,\frac{\Gamma\left(\n/2 - 2\right)}{32\pi^{\n/2}\left(s^2 + i\epsilon\right)^{\n/2 - 3}}\left(\frac{\mathcal{D}^{1/2}_{12}}{s^2 + i\epsilon} + \frac{1-6\xi}{12\left(\n - 6\right)}R\left(x_0\right)\right).
\end{equation}
At the Wilson-Fisher fixed point with conformal coupling, and at leading order in $\lambda$, the Ricci scalar term cancels and we obtain
\begin{equation}
\label{eq:C4WF}
\eval{\hat{C}_{\phi\phi}^{\phi^2}\left(\Delta x, x_0\right)}_{\rm WF} = \mathcal{D}^{1/2}_{12} s^{\gamma_{\phi^2}}.
\end{equation}
Notice that sometimes a different normalization is chosen for the operator $\phi^2$, in which one works with $\mathcal{O}_2= \phi^2/\sqrt{2}$, so that its two point function has unit coefficient (see e.g.~\cite{Rattazzi:2008pe}).

This result can be easily generalized to the ${\rm O}(N)$ Wilson-Fisher fixed point by considering a multiplet of massless scalar fields $\phi_i$ in the vector representation of ${\rm O}(N)$ interacting with a quartic interaction~\cite{Wilson:1972cf}. The fixed point is obtained for $(N+8)\lambda/96\pi^2 = \epsilon$. Also in this case, a similar analysis can be carried out for the trace and traceless parts of the $\phi_i^2$ operator. At leading order in $\epsilon$ the Ricci scalar term vanishes and one is left with a result similar to Eq.~\eqref{eq:C4WF} (up to different normalization factors).

\section{Scalar primaries in CFTs coupled to gravity}
\label{s:Conformal Primaries}

An interesting generalization of the above procedure is to OPE coefficients of scalar primary fields in arbitrary (strongly or weakly interacting) CFTs coupled to a curved background. We consider here theories that are conformal in flat space: their flat space two- and three-point functions, and hence their OPEs, are completely constrained by symmetries. The coupling to the gravitational background breaks conformal invariance in a calculable way, and the curved-space leading and subleading OPE coefficients: 
\begin{equation}
\label{eq:conformal OPE}
\lim_{x_{1, 2} \rightarrow x_0} \mathcal{O}_\Delta\left(x_1\right)\mathcal{O}_\Delta\left(x_2\right) = \hat{C}_{\mathcal{O}_1 \mathcal{O}_2}^{\mathbb{1}} \mathbb{1} + \hat{C}_{\mathcal{O}_1 \mathcal{O}_2}^{\mathcal{O}_0^\prime} \mathcal{O}_{\Delta^\prime}\left(x_0\right) + \dots,
\end{equation}
have universal gravitational corrections (we focus on well-behaved, Hadamard states).
We use the shorthand notation
\begin{equation}
\mathcal{O}_{\Delta}\left(x_i\right) = \mathcal{O}_i \quad \text{and} \quad \mathcal{O}_{\Delta^\prime}\left(x_i\right) = \mathcal{O}_i^\prime \, .
\end{equation}
To be clear, we consider local CFTs in \textit{flat} spacetime and couple them background gravitational metric through the stress-energy tensor. Finally, we note that, for simplicity, we will work in Euclidean space and so we will drop the time-ordered products and the $i\varepsilon$ terms. In this section, we will assume that the one-point functions vanish, $\expval{\mathcal{O}\left(x_0\right)}_g = 0$, a safe assumption since we will be working at linear order in $h$ in $d > 2$ and the expectation value of the stress tensor enters at higher orders in curvature through the conformal anomaly.

The method used in Ref.~\cite{Chavda:2025ijz} to compute the leading OPE coefficient was to evaluate the integral over the three-point function in Eq.~\eqref{eq:first_order} exactly. This was possible because the three-point function is completely fixed by conformal invariance. To compute the subleading OPE coefficient in this way, however, one would need a four-point function, which is only fixed up to conformal cross ratios (see e.g. \cite{Poland:2018epd}). Therefore, in this section we will frame the computation in a different way that solely relies on the knowledge of Ward identities. In computing the subleading OPE coefficient, we will use the method of regions, a strategy predominately used in the efficient computation of Feynman integrals \cite{Beneke:1997zp}.

\subsection{Leading OPE coefficient}
\label{s:leading}
To extract the leading OPE coefficient in Eq.~\eqref{eq:conformal OPE}, we simply need to compute the two-point function following Eq.~\eqref{eq:first_order}. The leading flat space contribution is
\begin{equation}
\expval{\mathcal{O}_1 \mathcal{O}_2}_0 = \frac{C_d}{x_{12}^{2\Delta}},
\end{equation}
where $C_d$ is a normalization prefactor, usually chosen to be one in the CFT conventions. We use $x_{12}^\mu = x_1^\mu - x_2^\mu$ in place of $\Delta x^\mu$ to avoid confusion, since $\Delta$ will denote the scaling dimension. The next-to-leading-order term is given by the integral
\begin{equation}
\label{eq:First Order Correction}
\expval{\mathcal{O}_1 \mathcal{O}_2}^{\left(1\right)}_g = \frac{1}{2} \int h_{\mu\nu}\left(y\right) \expval{\mathcal{O}_1 T^{\mu\nu}\left(y\right) \mathcal{O}_2}_0 \dd[d]{y}.
\end{equation}

To simplify the integral, we use the fact that the metric perturbation, $h_{\mu\nu}$, is a symmetric tensor on a Riemannian manifold and hence admits the following decomposition 
\begin{equation}
\label{eq:Metric Decomp}
h_{\mu\nu} = 2\omega \delta_{\mu\nu} + \partial_\mu \xi_\nu + \partial_\nu \xi_\mu + h_{\mu\nu}^{\text{TT}},
\end{equation}
where the first piece contains the trace, the second piece is the longitudinal term, and the last piece is traceless and transverse. To obtain explicit expressions for these components, we work for the moment in Riemann normal coordinates, where
\begin{equation}
h_{\mu\nu}\left(y\right) = -\frac{1}{3} R_{\mu\rho\nu\sigma}\left(x_0\right)\left(y - x_0\right)^\rho \left(y - x_0\right)^\sigma.
\end{equation}
We can restore full gauge dependence, which serves as a nontrivial check for our results since the final expressions must be gauge invariant. To do so, we will shift $\xi^\mu \rightarrow \xi^\mu + \zeta^\mu$ and check that the result is independent of $\zeta^\mu$. Matching the expression of $h_{\mu\nu}$ in Riemann normal coordinates to the decomposition, we find 
\begin{subequations}\label{eq:RNC}
\begin{align}
h^{\text{TT}}_{\mu\nu} &= -\frac{1}{3} C_{\mu\rho\nu\sigma} \hat{y}^\rho \hat{y}^\sigma, \\
\xi_\mu &= \frac{1}{3\left(d - 2\right)} R_{\rho\sigma} \hat{y}^\rho \hat{y}^\sigma \hat{y}_\mu - \frac{1}{6\left(d - 2\right)} R_{\mu\rho} \hat{y}^\rho \hat{y}^2 - \frac{R}{12\left(d - 1\right)\left(d - 2\right)}\hat{y}_\mu \hat{y}^2,\\
\omega &= -\frac{1}{2\left(d - 2\right)}\left(R_{\rho\sigma}\hat{y}^\rho \hat{y}^\sigma - \frac{R}{2\left(d - 1\right)} \hat{y}^2\right). 
\end{align}
\end{subequations}
where all curvature quantities are defined at the midpoint $x_0$ and $\hat{y} = y - x_0$. 

Applying this decomposition to Eq.~\eqref{eq:First Order Correction}, we find that there are three pieces. The transverse traceless piece is
\begin{equation}
\expval{\mathcal{O}_1 \mathcal{O}_2}^{\left(1\right)}_g = -\frac{1}{6} C_{\mu\rho\nu\sigma}\left(x_0\right) \int \hat{y}^\rho \hat{y}^\sigma \expval{\mathcal{O}_1 T^{\mu\nu}\left(y\right) \mathcal{O}_2}_0 \dd[d]{y}.
\end{equation}
Note that the integral, by Lorentz invariance, can only depend on $\delta^{\mu\nu}$ and $\Delta x^\mu$. Since the Weyl tensor is completely traceless, any contributions from $\delta^{\mu\nu}$ will automatically be zero. Therefore, at this order, the only potential contribution is of the form $C_{\mu\rho\nu\sigma} \Delta x^\mu \Delta x^\rho \Delta x^\nu \Delta x^\sigma$ which is automatically zero by the antisymmetry properties of the Weyl tensor. 

As a result, the first order integral in Eq.~\eqref{eq:First Order Correction} has only two contributions
\begin{equation}
\expval{\mathcal{O}_1 \mathcal{O}_2}^{\left(1\right)}_g = \int \omega\left(y\right) \expval{\mathcal{O}_1 T^\mu_{\hphantom{\mu}\mu}\left(y\right) \mathcal{O}_2}_0 \dd[d]{y} - \int \xi_\nu \partial_\mu \expval{\mathcal{O}_1 T^{\mu\nu}\left(y\right) \mathcal{O}_2}_0 \dd[d]{y}.
\end{equation}
In the second term, we integrated by parts and used symmetry of the stress-energy tensor to simplify. The correlators appearing in the two integrals are fixed by two flat space Ward identities \cite{Simmons-Duffin:2016gjk}:
\begin{subequations}
\begin{align}
\expval{\mathcal{O}_1 T^\mu_{\hphantom{\mu}\mu}\left(y\right) \mathcal{O}_2}_0 &= -\Delta\left(\delta^{\left(d\right)}\left(y - x_1\right) + \delta^{\left(d\right)}\left(y - x_2\right)\right)\expval{\mathcal{O}_1\mathcal{O}_2}_0,\\
\partial_\mu \expval{\mathcal{O}_1 T^{\mu\nu}\left(y\right) \mathcal{O}_2}_0 &= -\left(\delta^{\left(d\right)}\left(y - x_1\right)\partial_{x_1}^\nu + \delta^{\left(d\right)}\left(y - x_2\right)\partial_{x_2}^\nu\right)\expval{\mathcal{O}_1\mathcal{O}_2}_0.
\end{align}
\end{subequations}
Therefore, the integral becomes
\begin{equation}
\expval{\mathcal{O}_1 \mathcal{O}_2}^{\left(1\right)}_g = -\Delta\left[\omega\left(x_1\right) + \omega\left(x_2\right) + 2 \frac{x_{12}^\mu\left(\xi_1 - \xi_2\right)_\mu}{x_{12}^2} \right]\expval{\mathcal{O}_\Delta\left(x_1\right) \mathcal{O}_{\Delta^\prime}\left(x_2\right)}.
\end{equation}
Using the previous expressions for $\omega$ and $\xi$ in Riemann normal coordinates, Eq.~\eqref{eq:RNC}, 
\begin{subequations}
\begin{align}
\label{eq:omega RNC}
\omega\left(x_1\right) + \omega\left(x_2\right) &= -\frac{1}{8\Delta_0}\left(R_{\rho\sigma} x_{12}^\rho x_{12}^\sigma - \frac{R}{2\left(d - 1\right)} x_{12}^2\right),\\
\frac{x_{12}^\mu\left(\xi_1 - \xi_2\right)_\mu}{x_{12}^2} &= \frac{1}{48\Delta_0}\left(R_{\rho\sigma} x_{12}^\rho x_{12}^\sigma - \frac{R}{2\left(d - 1\right)} x_{12}^2\right).
\label{eq:xi RNC}
\end{align}
\end{subequations}
Therefore, the first order correction becomes
\begin{equation}
\expval{\mathcal{O}_1 \mathcal{O}_2}^{\left(1\right)}_g = \frac{\Delta}{12\Delta_0}\left(R_{\rho\sigma} x_{12}^\rho x_{12}^\sigma - \frac{R \, x_{12}^2}{2\left(d - 1\right)}\right)\frac{C_d}{x_{12}^{2\Delta}}.
\end{equation}

To obtain the full two-point function, we add the flat space contribution. As a non-trivial check, we also re-instate gauge dependence by a shift $\xi \mapsto \xi + \zeta$, to verify that the result can be expressed in gauge invariant form. We have:
\begin{equation}
\expval{\mathcal{O}_1 \mathcal{O}_2}_g = \left(1 + \frac{\Delta}{12\Delta_0} R_{\rho\sigma} x_{12}^\rho x_{12}^\sigma - \frac{\Delta R \, x_{12}^2}{24\left(d - 1\right)\Delta_0} - \frac{2\Delta x_{12}^\mu \left(\zeta_1 - \zeta_2\right)_\mu}{x_{12}^2}\right) \frac{C_d}{x_{12}^{2\Delta}} ,
\end{equation}
and using the expressions in Appendix~\ref{a:ImportantIdentities} for the geodesic distance and the van Vleck determinant we arrive at the generally covariant final form:  
\begin{equation}
\label{eq:conformal_2pt_final}
\expval{\mathcal{O}_\Delta\left(x_1\right) \mathcal{O}_{\Delta}\left(x_2\right)}_g = \frac{C_d}{s^{2\Delta}}\left(\mathcal{D}_{12}^{\Delta/\left(2\Delta_0\right)} - \frac{\Delta/\Delta_0}{24\left(d - 1\right)} s^2 R\left(x_0\right)\right).
\end{equation}
There are a number of checks for this result. First, the final answer is a geometric invariant: there is no residual gauge dependence on $\zeta^\mu$. Second, the result agrees with that obtained by direct integration of the full three-point function, as first derived in Ref.~\cite{Chavda:2025ijz}. Moreover, upon setting $\Delta = \Delta_0$, we find that this correctly reduces to the two-point function of two Lorentz scalars with conformal coupling to the Ricci scalar, as given in Eq.~\eqref{eq:nonmin_GF}. Finally, in the special case of a conformally flat background, the two-point function can be related to the flat space two-point correlator up to Weyl rescalings, as we verify in Appendix~\ref{a:Weyl}.   

\subsection{Subleading OPE coefficient}

We now extend this method to the subleading OPE coefficient in Eq.~\eqref{eq:conformal OPE}, merging two conformal primaries $\mathcal{O}$ into a new conformal primary $\mathcal{O}^\prime$. If we then expand Eq.~\eqref{eq:conformal OPE} to first order in metric perturbations, following Eq.~\eqref{eq:OPE Grav Expansion}, we would find
\begin{equation}
\label{eq:Expectation}
\lim_{x_1 \rightarrow x_2} \expval{\mathcal{O}_1 \mathcal{O}_2 \mathcal{O}_3^\prime}_g = \hat{C}_{\mathcal{O}_1 \mathcal{O}_2}^{\mathcal{O}_3'\: (0)} \hat{C}_{\mathcal{O}_0^\prime \mathcal{O}_3^\prime}^{\mathbb{1} \: (0)} + \hat{C}_{\mathcal{O}_1 \mathcal{O}_2}^{\mathcal{O}_3'\:(0)} \hat{C}_{\mathcal{O}_0^\prime \mathcal{O}_3^\prime}^{\mathbb{1}\:{(1)}} + \hat{C}_{\mathcal{O}_1 \mathcal{O}_2}^{\mathcal{O}_3'\:(1)} \hat{C}_{\mathcal{O}_0^\prime \mathcal{O}_3^\prime}^{\mathbb{1} \: {(0)}} + \dots \, .
\end{equation}
The first term is the flat space term, the second is the correction to the two-point function which was computed in the previous subsection, and the third term is what we are currently interested in, the term involving the correction to the three-point OPE coefficient. The flat space terms can be determined from knowledge of the full three-point function 
\begin{equation}
\label{eq:3PointFunction}
\expval{\mathcal{O}_\Delta\left(x_1\right) \mathcal{O}_\Delta\left(x_2\right) \mathcal{O}_{\Delta'}\left(x_3\right)}_0 = \frac{\g \, C_d}{\abs{x_1 - x_2}^{2\Delta - \Delta'} \abs{x_1 - x_3}^{\Delta'} \abs{x_2 - x_3}^{\Delta'}}.
\end{equation}
In particular, expanding this in the limit $x_{1, 2} \rightarrow x_0$, and factorizing the two-point function we can extract the following flat space OPE factors:
\begin{subequations}
\begin{align}
\label{eq:flatOPE1}
C_{\mathcal{O}_1 \mathcal{O}_2}^{\mathcal{O}_0^\prime} &= \frac{\g}{x_{12}^{2\Delta - \Delta^\prime}},\\
C_{\mathcal{O}_1 \mathcal{O}_2}^{\partial \partial \mathcal{O}_0^\prime} &= \frac{\g}{x_{12}^{2\Delta - \Delta^\prime}} \left(\frac{1}{8\left(\Delta^\prime + 1\right)} x_{12}^\mu x_{12}^\nu + \frac{\Delta^\prime}{16\left(\Delta^\prime + 1\right)\left(\Delta_0 - \Delta^\prime\right)} x_{12}^2 \delta^{\mu\nu}\right).
\label{eq:flatOPE2}
\end{align}
\end{subequations}
The second corresponds to the coefficient of a descendant operator, but as we will see this will contribute to the gravitational corrections of the first OPE coefficient. Intuitively, two derivatives contribute at the same order as curvature. Hence, one needs to carefully account for flat space descendant contributions to the gravitationally dressed coefficients of a primary. 

The first order gravitational correction to the three-point function, from which one can determine the terms in Eq.~\eqref{eq:Expectation}, is given by the insertion of a stress-energy tensor: 
\begin{equation}
\label{eq:MainIntegral}
\expval{\mathcal{O}_1 \mathcal{O}_2 \mathcal{O}_3^\prime}_g = \expval{\mathcal{O}_1 \mathcal{O}_2 \mathcal{O}_3^\prime}_0 + \frac{1}{2} \int h_{\mu\nu}\left(y\right) \expval{\mathcal{O}_1 \mathcal{O}_2 T^{\mu\nu}\left(y\right) \mathcal{O}_3^\prime}_0 \dd[d]{y}.
\end{equation}
As in Section~\ref{s:intscalars}, we will consider a limit in which $x_1, x_2 \mapsto x_0$ first and then $x_3 \mapsto x_0$. Unlike earlier, however, we no longer have momentum space diagrams that separately contribute to each term of Eq.~\eqref{eq:Expectation}.

\begin{figure}
\begin{center}
\begin{tikzpicture}
\filldraw[yellow!15] (0, 0) circle (1.5cm);
\draw[yellow, line width=0.4mm] (0, 0) circle (1.5cm);
\draw[orange, line width=0.3mm] (0, 0) -- node[anchor=south, xshift=-0.1cm] {$R$} (1.5*0.707, 1.5*0.707);
\filldraw[black] (-0.8*0.707, 0.8*0.707) circle (0.05cm);
\node at (-0.8*0.707 - 0.28, 0.8*0.707) {$x_1$};
\filldraw[black] (0.8*0.5, -0.8*0.866) circle (0.05cm);
\node at (0.8*0.5 + 0.28, -0.8*0.866) {$x_2$};
\filldraw[black] (0, 0) circle (0.05cm);
\node at (-0.28, 0) {$x_0$};
\filldraw[black] (2.25*0.866, 2.25*0.5) circle (0.05cm);
\node at (2.25*0.866 + 0.28, 2.25*0.5) {$x_3$};
\node[orange] at (0, -1.2) {$\mathcal{B}$};
\end{tikzpicture}
\end{center}
\caption{The Regions. In our analysis, we take the OPE of $\mathcal{O}_1$ and $\mathcal{O}_2$. Inside the integral, the stress-energy tensor insertion can be either inside or outside the ball $\mathcal{B}$, which defines the area in which one can take the OPE of $\mathcal{O}_1$ and $\mathcal{O}_2$. This splits up the integral into two pieces: one which contributes to the Hadamard tail and one which contributes to the OPE correction. The size of $\mathcal{B}$ is set by any $R$ which is large enough to encompass $x_1$ and $x_2$ but does not contain $x_3$.}
\label{fig:Regions}
\end{figure}
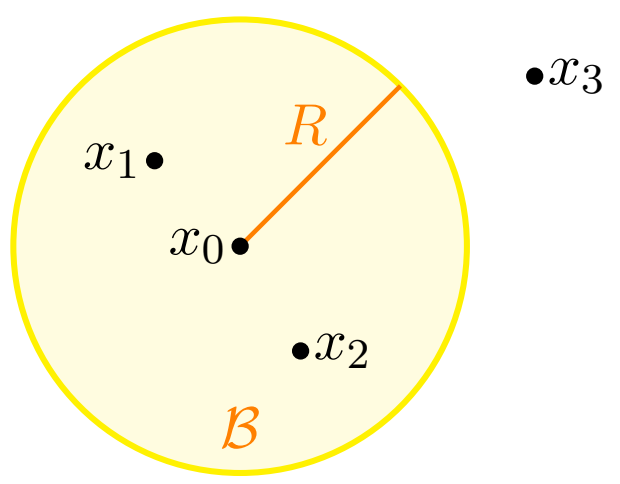

To proceed, we instead note that the integral in Eq.~\eqref{eq:MainIntegral} is dominated by two parametrically different regimes for the stress-tensor insertion, which we treat with the method of regions \cite{Beneke:1997zp}. The two regions are
\begin{equation}
\text{(i)}\quad |y - x_0| \sim |x_3 - x_0| \gg x_{12}\ , \qquad\qquad
\text{(ii)}\quad |y - x_0| \sim x_{12} \ll |x_3 - x_0|\ ,
\end{equation}
illustrated in Figure~\ref{fig:Regions}. In each region we expand the integrand in the corresponding small ratio and then integrate over all of spacetime. As a result, there will be overlap integrals. Much work has gone into studying these overlap integrals in the study of Feynman integrals. For example, when using a sector decomposition, such overlap integrals are scaleless and hence zero in dimensional regularization \cite{Heinrich:2021dbf}. 

In region (i), the stress tensor sits far from the pair $\mathcal{O}_1\mathcal{O}_2$, so we may fuse the pair first,
\begin{equation}
\mathcal{O}_1 \mathcal{O}_2 = \hat{C}_{\mathcal{O}_1\mathcal{O}_2}^{\mathbb{1}}\,\mathbb{1} + \hat{C}_{\mathcal{O}_1\mathcal{O}_2}^{\mathcal{O}'_0}\,\mathcal{O}'_0 + \dots,
\end{equation}
where the expansion is about the midpoint $x_0$ and the ellipsis denotes descendants and higher-dimension primaries. The surviving object is the flat-space OPE coefficient multiplying a three-point function fixed by conformal symmetry,
\begin{equation}
\label{eq:RegionOne}
\begin{aligned}
\eval{\expval{\mathcal{O}_1 \mathcal{O}_2 \mathcal{O}^\prime_3 T^{\mu\nu}(y)}}_{\rm (i)}
&= C_{\mathcal{O}_1\mathcal{O}_2}^{\mathcal{O}^\prime_0}(x_1,x_2)\,
\expval{\mathcal{O}^\prime_0 \mathcal{O}^\prime_3 T^{\mu\nu}(y)} \\
&\quad + \left(C_{\mathcal{O}_1 \mathcal{O}_2}^{\partial \partial \mathcal{O}_0^\prime}\right)^{\rho\sigma} \eval{\langle \partial_\rho\partial_\sigma \mathcal{O}_z^\prime \mathcal{O}_3^\prime T^{\mu\nu}\left(y\right)\rangle}_{z = x_0} + \dots,
\end{aligned}
\end{equation}
where we used the fact that the identity term drops out because $\expval{\mathcal{O}^\prime_3 T^{\mu\nu}(y)} = 0$. We also include the first descendant term since, as discussed before, it will contribute to the desired order.

In region (ii), the stress tensor lies within the same neighborhood of $x_0$ as $\mathcal{O}_1$ and $\mathcal{O}_2$, and must be fused together with them,
\begin{equation}
\label{eq:RegionTwo}
\eval{\expval{\mathcal{O}_1 \mathcal{O}_2 \mathcal{O}'_3 T^{\mu\nu}(y)}}_{\rm (ii)}
= \left(C_{\mathcal{O}_1 \mathcal{O}_2 T_y}^{\mathcal{O}^\prime_0}\right)^{\mu\nu}
\expval{\mathcal{O}^\prime_0 \mathcal{O}'_3} + \dots,
\end{equation}
where again the identity term drops out, now because $\expval{\mathcal{O}'_3} = 0$. The two regions therefore separately contribute to the second and third terms anticipated in Eq.~\eqref{eq:Expectation}.

The takeaway is that region (i) dresses the flat-space OPE coefficient with the gravitational correction to the two-point function computed in the previous subsection, while region (ii) supplies the correction to the OPE coefficient itself, which is the object we are after. To make the integrals easy to evaluate, we extend the integration regions to cover the entirety of spacetime. To compensate, we have to subtract off overlap integrals. These overlap integrals are Eq.~\eqref{eq:RegionOne} integrated in region (ii) and Eq.~\eqref{eq:RegionTwo} integrated in region (i). In these new regions, we can expand a second time, giving the same integrand in regions (i) and (ii) since the expansion commute. The integrals can be added together, giving the final line below where we match to Eq.~\eqref{eq:Expectation}:
\begin{subequations}
\begin{align}
\hat{C}_{\mathcal{O}_0^\prime \mathcal{O}_3^\prime}^{\mathbb{1}\:{(1)}} &= \frac{1}{2}\int h_{\mu\nu}\left(y\right) \expval{\mathcal{O}_0^\prime \mathcal{O}_3^\prime T^{\mu\nu}\left(y\right)}_0 \dd[d]{y},\\ 
\hat{C}_{\mathcal{O}_1 \mathcal{O}_2}^{\mathcal{O}_3'\:(1)} &= \frac{1}{2} \int h_{\mu\nu}\left(y\right) \left(C^{\mathcal{O}_0^\prime}_{\mathcal{O}_1 \mathcal{O}_2 T_y}\right)^{\mu\nu} \dd[d]{y} \\
&\quad - \frac{1}{2} \int h_{\mu\nu}\left(y\right)\left[C_{\mathcal{O}_1 \mathcal{O}_2}^{\mathcal{O}_0^\prime} \left(C_{\mathcal{O}_0^\prime T_y}^{\mathcal{O}_0^\prime}\right)^{\mu\nu} + \left(C_{\mathcal{O}_1 \mathcal{O}_2}^{\partial \partial \mathcal{O}_0^\prime}\right)^{\rho\sigma} \partial_\rho \partial_\sigma \eval{\left(C_{\mathcal{O}_z^\prime T_y}^{\mathcal{O}_0^\prime}\right)^{\mu\nu} }_{z = x_0}  \right] \dd[d]{y}, \nonumber
\label{eq:3ptCFTcoeff}
\end{align}
\end{subequations}
where all coefficients on the RHS are flat space ones.
The first of these equations coincides with the computation already discussed in the Section~\ref{s:leading}, while the second contributes to the result we are after.

The first term in the last line is scaleless since it can only depend on $y$ and hence vanishes in dimensional regularization. As mentioned earlier, this is a typical feature encountered when using the method of regions \cite{Heinrich:2021dbf}. The second term, however, has more than one scale and does not vanish. Therefore, the first order correction to the OPE coefficient we are after is: 
\begin{equation}
\label{eq:3pt Main Integral}
\hat{C}_{\mathcal{O}_1 \mathcal{O}_2}^{\mathcal{O}_3'\:(1)} = \frac{1}{2} \int h_{\mu\nu}\left(y\right) \left[\left(C^{\mathcal{O}_0^\prime}_{\mathcal{O}_1 \mathcal{O}_2 T_y}\right)^{\mu\nu} - \left(C_{\mathcal{O}_1 \mathcal{O}_2}^{\partial \partial \mathcal{O}_0^\prime}\right)^{\rho\sigma} \partial_\rho \partial_\sigma \eval{\left(C_{\mathcal{O}_z^\prime T_y}^{\mathcal{O}_0^\prime}\right)^{\mu\nu} }_{z = x_0} \right] \, .
\end{equation}
We now implement the same procedure used in the previous section. It is important to first write down the following Ward identities (upon integration by parts):
\begin{subequations}
\begin{align}
\left(C_{\mathcal{O}_1 \mathcal{O}_2 T_y}^{\mathcal{O}_0^\prime}\right)^\mu_{\hphantom{\mu}\mu} &= -\Delta\left[\delta^{\left(d\right)}\left(y - x_1\right) + \delta^{\left(d\right)}\left(y - x_2\right)\right] C_{\mathcal{O}_1 \mathcal{O}_2}^{\mathcal{O}_0^\prime},\\
\partial_\mu \left(C_{\mathcal{O}_1 \mathcal{O}_2 T_y}^{\mathcal{O}_0^\prime}\right)^{\mu\nu} &= \left(2\Delta - \Delta^\prime\right)\frac{x_{12}^\nu}{x_{12}^2}\left[\delta^{\left(d\right)}\left(y - x_1\right) - \delta^{\left(d\right)}\left(y - x_2\right)\right] C_{\mathcal{O}_1 \mathcal{O}_2}^{\mathcal{O}_0^\prime}.
\end{align}
\end{subequations}
These can be determined directly from the Ward identities on correlation functions, by writing down the OPE expansion of both sides. Using the decomposition~\eqref{eq:Metric Decomp} and the Ward identities in Eq.~\eqref{eq:3pt Main Integral}, we find 
\begin{equation}
\hat{C}_{\mathcal{O}_1 \mathcal{O}_2}^{\mathcal{O}_3'\:(1)} = \left[-\Delta\left(\omega_1 + \omega_2\right) - \left(2\Delta-\Delta^\prime\right) \frac{x_{12} \cdot \xi_{12}}{x_{12}^2}\right] C_{\mathcal{O}_1 \mathcal{O}_2}^{\mathcal{O}_0^\prime} + \Delta^\prime \left(C_{\mathcal{O}_1 \mathcal{O}_2}^{\partial \partial \mathcal{O}_0^\prime}\right)^{\rho\sigma} \partial_\rho \partial_\sigma \omega_0,
\end{equation}
where we work up to second order in derivatives. This is also automatically gauge invariant since, in the first term, it goes as $x_{12}^{-2\Delta + \Delta^\prime}$ so changing that to a general geodesic distance would exactly cancel a shift $\xi \rightarrow \xi + \zeta$. All that remains is to compute the OPE coefficients.

Using the expressions for $\omega$ and $\xi$---Eqs.~\eqref{eq:omega RNC} and~\eqref{eq:xi RNC}---and the flat space OPE coefficients---Eqs.~\eqref{eq:flatOPE1} and~\eqref{eq:flatOPE2}---we arrive at 
\begin{equation}
\begin{aligned}
\hat{C}_{\mathcal{O}_1 \mathcal{O}_2}^{\mathcal{O}_3'\:(1)} &= \frac{1}{12\Delta_0}\left(\Delta + \frac{\Delta^\prime\left(\Delta^\prime - 2\right)}{4\left(\Delta^\prime + 1\right)}\right)\left(R_{\rho\sigma} x_{12}^\rho x_{12}^\sigma - \frac{R}{2\left(d - 1\right)}s^2\right) \frac{\g}{s^{2\Delta - \Delta^\prime}}\\
&\quad - \frac{R}{2\left(d - 1\right)}\left(\frac{\left(\Delta^\prime\right)^2}{16\left(\Delta^\prime + 1\right)\left(\Delta_0 - \Delta^\prime\right)} s^2\right)\frac{\g}{s^{2\Delta - \Delta^\prime}}. 
\end{aligned}
\end{equation}
Including the flat space piece and covariantizing, we arrive at the final result
\begin{equation}
\label{eq:conformal_3pt_final}
\hat{C}_{\mathcal{O}_1 \mathcal{O}_2}^{\mathcal{O}_3^\prime} = \frac{\g}{s^{2\Delta - \Delta^\prime}}\left(\mathcal{D}_{12}^{a/\left(2\Delta_0\right)} - \frac{b \, s^2}{2\left(d - 1\right)} R\right),
\end{equation}
where
\begin{subequations}
\begin{align}
a &= \Delta + \frac{\Delta^\prime \left(\Delta^\prime - 2\right)}{4\left(\Delta^\prime + 1\right)},\\
b &= \frac{a}{12\Delta_0} + \frac{\left(\Delta^\prime\right)^2}{16\left(\Delta^\prime + 1\right)\left(\Delta_0 - \Delta^\prime\right)}.
\end{align}
\end{subequations}
We can check this expression in several ways. First, substituting the scaling dimensions of the Wilson--Fischer fixed point from Eqs.~\eqref{eq:WF Deltaphi} and~\eqref{eq:WF Deltaphisq}, we find $a=1/2$ and $b=0$, matching the independent result in Eq.~\eqref{eq:C4WF}. Second, our expression reduces to the conformally flat background result detailed in Appendix~\ref{a:Weyl}. The derivation for a conformally flat background was presented independently in Ref.~\cite{Das:2026tjz}. We have verified that our result, upon expanding in curvature, agrees with Ref.~\cite{Das:2026tjz} for $\Delta_1 = \Delta_2 = \Delta$.

\section{Smearing-in-time in curved spacetime}
\label{s:algebra}

With explicit short-distance expansions at our disposal, we are able to check explicitly the validity of the smearing-in-time procedure used in defining the algebra of observables in studies of quantum field theory in curved spacetime with an observer, following work of Witten~\cite{Witten:2023qsv}.  
By “algebra of observables,” we mean the set of all Hermitian operators on a Hilbert space within a bounded region of spacetime, typically subject to additional physically motivated constraints such as microcausality. Concrete axiomatic formulations of this framework can be found in \cite[Chapter 1.1]{Brunetti:2015vmh} or \cite{Hollands:2009bke}.

In an asymptotically flat spacetime, the observer can be placed at infinity and so the energy of the observer does not significantly back-react on the geometry. By contrast, in a general curved background, such as the static patch of de Sitter spacetime, there may be no suitable asymptotic region in which the observer decouples. In that case, one must include the observers' gravitational effects in defining an algebra of observables. Nevertheless, in a local neighborhood, the equivalence principle ensures the validity of a weak-field approximation and one can construct the algebra of observables around a point on the observer's worldline. We will use a scalar field as an example and follow the smearing procedure outlined in Ref.~\cite{Witten:2023qsv}.

Well-defined Hilbert-space operators acting on the vacuum, or other physical states, must have finite norms. In contrast, local quantum fields such as the scalar $\phi\left(x\right)$ are operator-valued distributions and their product is plagued by UV divergences when multiple fields are evaluated at the same spacetime point. For example, given the short-distance expansion of two fields in Eq.~\eqref{eq:2ptHadamard},
\begin{equation}
\abs{\phi\left(x\right)\ket{0}}^2 = \mel{0}{\phi\left(x\right)\phi\left(x\right)}{0} \to \infty.
\end{equation}
With that said, local quantum fields can be used to generate well-defined operators. Typically, one smears these fields against suitable test functions of space and time, thereby integrating over an extended region rather than evaluating them at a single point. As pointed out in Ref.~\cite{Witten:2023qsv}, spatial smearing is not sufficient to have a well-defined operator, which requires smearing in real (Lorentzian) time. As we will now show, the time-smeared field $\phi_f$, defined~as 
\begin{equation}
\phi_f\left(\vb{x}_0\right) \equiv \int \phi\left(t, \vb{x}_0\right) f\left(t\right) \dd{t},
\end{equation}
where $f\left(t\right)$ is a test function which is assumed to be smooth and compactly supported, is a proper element of the algebra of observables: a~well-defined operator. To see this, one has to show that the norm of the operator acting on the vacuum, 
\begin{equation}\label{eq:norm}
\abs{\phi_f\left(\vb{x}_0\right)\ket{0}}^2 = \int \phi\left(t, \vb{x}_0\right) \phi\left(t', \vb{x}_0\right) f\left(t\right) f^\ast\left(t'\right) \dd{t} \dd{t}' \, ,
\end{equation}
is finite. Let us discuss first the case of weakly coupled fields. Using the results of Section~\ref{s:intscalars} and Eq.~\eqref{eq:2ptHadamard_OPE} in $d = 4$, the finiteness of the above norm is equivalent to convergence of the integrals
\begin{equation}
\label{eq:HadamardAlgChecks}
\int \frac{1}{\left(t - t' + i\epsilon\right)^2} f\left(t\right) f^\ast\left(t'\right) \dd{t}\dd{t'} \quad \text{and} \quad \int \log\left(t - t' + i\epsilon\right) f\left(t\right) f^\ast\left(t'\right) \dd{t}\dd{t'}.
\end{equation}
Notice that the two fields in eq.~\eqref{eq:norm} have been both evaluated at $\vb{x}_0$ and as a consequence the spatial dependence drops out. The convergence of these integrals therefore shows that temporal smearing is sufficient in defining the finite-norm operator $\phi_f\left(\vb{x}_0\right)$.

The integral on the left can be shown to be convergent using \cite{Witten:2023qsv}, 
\begin{equation}
\label{eq:IBPpower}
\frac{1}{\left(t - t' + i\epsilon\right)^2} = C_n \partial_t^n \left(\left(t - t' + i\epsilon\right)^{n - 2} \log\left(t - t' + i\epsilon\right)\right),
\end{equation}
for sufficiently high $n$, with $C_n$ a constant.
Upon integrating by parts $n$ times, the remaining function is indeed convergent in the $\epsilon \rightarrow 0$ limit. 
The integral on the right of Eq.~\eqref{eq:HadamardAlgChecks}, on the other hand, is immediately convergent in the $\epsilon \rightarrow 0$ limit, since the logarithm itself is locally integrable near $t \to t'$.
This integration-by-parts argument works because these short-distance singularities remain bounded by a power law and logarithms. This power-law bound is precisely why one-dimensional (time) smearing, rather than a full four-dimensional smearing, is sufficient to define well-behaved operator insertions.

More generally, the condition that an algebra of observables is constructible in a local region near the observer's worldline is that the short-distance behavior of the operator, given by the OPE expansion function, is integrable. This condition holds for all physical theories in Minkowski space and is claimed by \cite{Witten:2023qsv} to be a way of generating an algebra of observables in a general curved spacetime. On the one hand, it is not obvious that the smearing-in-time procedure should carry over to curved spacetime where there is no preferred vacuum and so no preferred construction of the Hilbert space. This would imply that the operators as defined above cannot be assumed to be operators on a unique Hilbert space \cite{Hollands:2009bke}. On the other hand, the result should not be so surprising in the sense that the condition for generating a well-defined algebra of observables near an observers' worldline is based on short-distance physics, which is Minkowski-like by the equivalence principle \cite{Witten:2023qsv}. While the OPEs in curved spacetime are more complicated and include curvature-dependent terms, the short-distance behavior is not significantly more complicated. 

Going beyond weakly coupled two-point functions, all OPE singularities studied in the present paper, including those of strongly coupled CFTs in a curved background, take the form of $\left(t - t' + i\epsilon\right)^{-2b} \log^a\left(t - t' + i\epsilon\right)$ (e.g. in the four-point function, Eq.~\eqref{eq:phi4d4}, $b = -1/2$ and $a = 1$). Hence, the ability to define an algebra in a local region is equivalent to the convergence of the integral
\begin{equation}
\int \frac{\log^a\left(t - t' + i\epsilon\right)}{\left(t - t' + i\epsilon\right)^{2b}} f\left(t\right)f^\ast\left(t'\right) \dd{t} \dd{t'}. 
\end{equation}
Using the same logic, one can express the short distance singular structure for sufficiently high $n$ as 
\begin{equation}
\frac{\log^a\left(t - t' + i\epsilon\right)}{\left(t - t' + i\epsilon\right)^{2b}} = \partial_t^n\left[\sum_{k = 0}^{a} c_{k, n} \left(t - t' + i\epsilon\right)^{n - 2b} \log^{k + 1}\left(t - t' + i\epsilon\right)\right],
\end{equation}
where one can fix $a - 1$ of the $a$ $c_{k, n}$s to isolate the proper form on the right-hand side. The remaining $c_{k, n}$ is an overall constant. Upon integration by parts, the integral becomes manifestly convergent as $\epsilon \rightarrow 0$. Notice that the singular structure is milder than that of a power law, such as in Eq.~\eqref{eq:IBPpower} so well-defined operator insertions make the integrals considered more convergent. Additionally notice that the curvature terms are evaluated at constants points, in this instance on the observer's worldline, and so will not introduce any other singularities. Therefore, in addition to having a finite norm, products of real scalar fields can be used to generate elements of the algebra of observables using the suitably time-smeared operator $\phi_f\left(\vb{x}_0\right)$.

\section{Discussion}

Background-independent results for quantum field theory in curved spacetime are rare and often limited to weakly coupled field theories. Building up on the approach of Refs.~\cite{Komissarov:2022gax,Chavda:2025ijz} we have derived the leading universal curvature corrections to curved space OPE coefficients of interacting QFTs in a gravitational background. 
We have first done so for weakly coupled scalar fields in theories with cubic and quartic interactions, Eqs.~\eqref{eq:FinalCubicResult} and~\eqref{eq:C4minimal} respectively, including also nonminimal coupling for the latter, Eq.~\eqref{eq:C4nonminimal}. 
We have then considered the case of scalar primary operators of $d$-dimensional Conformal Field Theories coupled to general gravitational backgrounds, obtaining the closed-form expression Eq.~\eqref{eq:conformal_3pt_final}.
As an application, we have discussed and verified explicitly the validity of the smearing-in-time procedure to define the algebra of observables of an observer in curved spacetime, as put forward by Witten in Ref.~\cite{Witten:2023qsv}.

In the future, it would be interesting to generalize the analysis of interacting OPE coefficients to spinning fields, extending the vector and graviton two-point function results of Ref.~\cite{Chavda:2025ijz}. The study of correlators in a curved background, and graviton correlators in particular, could also find applications in ongoing efforts to obtain analytic results in gravitational classical self-force computations~\cite{Pound:2015tma,Kosmopoulos:2023bwc,Cheung:2023lnj}. Finally, it would be interesting to apply these technical results to problems of physical relevance, such as correlation functions in inflationary cosmology or the study of black holes and their perturbations. We leave these explorations to future works.

\acknowledgments
A.C.~is supported by the National Science Foundation Graduate Research Fellowship under Grant No. DGE 2437839. J.S. is supported by funds provided by the Center for Particle Cosmology at the University of Pennsylvania. This work is also partially supported by the U.S.~Department of Energy (award no.~DE-SC0011941). 
The authors acknowledge the use of AI tools to assist with code development.

\appendix

\section{Useful Identities}
\label{a:ImportantIdentities}

Throughout the text, we will use the following inverse Fourier transform,
\begin{align}
\label{eq:GenericI}
I_\alpha\left(x\right) &\equiv \int \frac{-i}{\left(k^2 - i\epsilon\right)^{\alpha}} e^{i k\cdot x} \frac{\dd[\n]{k}}{\left(2\pi\right)^\n},\\ \label{eq:GenericIresult}
&= \frac{\Gamma\left(\n/2 - \alpha\right)}{4^\alpha \pi^{\n/2}\Gamma\left(\alpha\right)\left(x^2 + i\epsilon\right)^{\n/2 - \alpha}} \; .
\end{align}
For large enough $\alpha$, $\alpha \geq d/2$, the integral is IR-divergent and is addressed in dimensional regularization.

Since the OPE coefficients we study in the main text depend on two spacetime points, their final form can be expressed in terms of \textit{bitensors}. For a review, see Refs.~\cite{DeWitt:1960fc, Poisson:2011nh}. For our purposes, it is important to know their expansions around flat space. For example, the geodesic distance is 
\begin{equation}
\label{eq:SigmaExp}
s^2\left(x_1, x_2\right) = \Delta x^2 + h_{\mu\nu}\left(x_0\right) \Delta x^\mu \Delta x^\nu + \frac{1}{24} \Delta x^\mu \Delta x^\nu \Delta x^\rho \Delta x^\sigma \partial_\rho \partial_\sigma h_{\mu\nu}\left(x_0\right) + \dots,
\end{equation}
where $\Delta x^\mu = x_1^\mu - x_2^\mu$ and $x_0 = \left(x_1 + x_2\right)/2$ is the midpoint. A second example of a biscalar is the van Vleck determinant $\mathcal{D}\left(x_1, x_2\right) \equiv \mathcal{D}_{12}$, which determines the extent to which geodesics focus or defocus \cite{VanVleck:1928zz, Visser:1992pz}. Its expansion about the midpoint, $x_0$, is 
\begin{equation}
\label{eq:vvexpansion}
\mathcal{D}_{12} = 1 + \frac{1}{6}R_{\mu\nu}\left(x_0\right)\Delta x^\mu \Delta x^\nu + \dots,
\end{equation}
whose derivation is given in Ref.~\cite{Poisson:2011nh}. 

\section{Conformal OPEs from Weyl transformations}
\label{a:Weyl}

In this appendix, we demonstrate that the OPE coefficients obtained in Section~\ref{s:Conformal Primaries} are consistent with knowledge of the full correlation functions on a conformally flat background, 
\begin{equation}
g_{\mu\nu} = e^{2\omega\left(x\right)} \delta_{\mu\nu},
\end{equation}
where we work in Euclidean space for simplicity. A conformal primary of scaling dimension~$\Delta$ is related to its flat space counterpart by a Weyl rescaling, 
\begin{equation}
\label{eq:weyl_map}
\mathcal O_g(x) = e^{-\Delta \omega(x)} \mathcal O_\delta (x).
\end{equation}
where $\delta$ denotes flat Euclidean space. Correlators in a conformally flat background therefore follow from flat space ones without any perturbative integral. As in the main text, we work with local inertial coordinates $\omega(x_0)=\partial_\mu\omega(x_0)=0$, and at first order in $\omega$. In this case $h_{\mu\nu}=2\omega\delta_{\mu\nu}$, so that
\begin{subequations}
\begin{align}
\label{eq:omega_to_R_1}
x_{12}^\mu x_{12}^\nu\partial_\mu\partial_\nu\omega(x_0) &= -\frac{1}{2\Delta_0}\left(R_{\mu\nu}\left(x_0\right)x_{12}^\mu x_{12}^\nu-\frac{x_{12}^2}{2(d-1)}R\left(x_0\right)\right),\\
\Box\omega(x_0) &= -\frac{R\left(x_0\right)}{2(d-1)}\,.
\label{eq:omega_to_R_2}
\end{align}
\end{subequations}
Moreover, from Eq.~\eqref{eq:SigmaExp}, 
\begin{equation}
s^2 = x_{12}^2 \left(1 + \frac{1}{12} x_{12}^\mu x_{12}^\nu \partial_\mu \partial_\nu \omega\left(x_0\right)\right).
\end{equation}

Considering now the two point function, we apply Eq.~\eqref{eq:weyl_map} to both operators and find 
\begin{equation}
\expval{\mathcal{O}_1 \mathcal{O}_2}_g = e^{-\Delta \omega_1 - \Delta \omega_2} \frac{C_d}{x_{12}^{2\Delta}}.
\end{equation}
Expanding the exponential and using 
\begin{equation}
\omega(x_1)+\omega(x_2) = \frac{1}{4} x_{12}^\mu x_{12}^\nu\partial_\mu\partial_\nu\omega(x_0)+\ldots \, ,
\end{equation}
and matching to the generally covariant expressions, we recover 
\begin{equation}
\langle\mathcal{O}_1\mathcal{O}_2\rangle_g
=\frac{C_d}{s^{2\Delta}}\left(D_{12}^{\Delta/(2\Delta_0)}-\frac{\Delta/\Delta_0}{24(d-1)}\,s^2R\right).
\end{equation}
This is in agreement with the two-point function found in Eq.~\eqref{eq:conformal_2pt_final}.

The three point function, however, introduces a subtlety, namely that derivatives do not commute with Weyl rescaling. Differentiating $\mathcal{O}^\prime_\delta=e^{\Delta^\prime\omega}\mathcal{O}^\prime_g$ twice and evaluating at $x_0$, one finds:
\begin{equation}
\partial_\mu\partial_\nu\mathcal{O}^\prime_\delta(x_0) = \partial_\mu\partial_\nu\mathcal{O}^\prime_g(x_0)+\Delta'\,\partial_\mu\partial_\nu\omega(x_0)\,\mathcal{O}^\prime_g(x_0)\, ,
\label{eq:descendant_mixing}
\end{equation}
where we used that the terms with a single derivative of $\omega$ vanish. 
The flat-space derivative term thus contains a piece proportional to $\mathcal{O}^\prime$ itself. To be consistent, we define the leading descendant to be derivatives of $\mathcal{O}_g$. If we instead used derivatives of $\mathcal{O}_\delta$, the extra $\omega$ term would be dropped. Moreover, at the needed order, the flat-space OPE about the midpoint reads
\begin{equation}
\mathcal{O}_{1,\delta}\,\mathcal{O}_{2,\delta}\supset C^{\mathcal{O}^\prime}_{\mathcal{O}_1 \mathcal{O}_2} \,\mathcal{O}^\prime_\delta(x_0)+\left(C_{\mathcal{O}_1 \mathcal{O}_2}^{\partial\partial\mathcal{O}^\prime}\right)^{\mu\nu}\partial_\mu\partial_\nu\mathcal{O}^\prime_\delta(x_0)\,,
\end{equation}
where the flat space OPE coefficients are given in Eq.~\eqref{eq:flatOPE1} and Eq.~\eqref{eq:flatOPE2}.

Using Eq.~\eqref{eq:weyl_map} on $\mathcal{O}_1$ and $\mathcal{O}_2$, and Eq.~\eqref{eq:descendant_mixing} on the derivative term, we find
\begin{subequations}
\begin{align}
\mathcal{O}_{1, g} \mathcal{O}_{2, g} &= e^{-\Delta\omega_1 - \Delta \omega_2} \mathcal{O}_{1, \delta} \mathcal{O}_{2, \delta},\\
&= e^{-\Delta\omega_1 - \Delta\omega_2} \bigg[\left(C^{\mathcal{O}^\prime}_{\mathcal{O}_1 \mathcal{O}_2} + \Delta^\prime \left(C^{\partial \partial \mathcal{O}^\prime}_{\mathcal{O}_1 \mathcal{O}_2}\right)^{\mu\nu} \partial_\mu \partial_\nu \omega\left(x_0\right)\right) \mathcal{O}^\prime_g\left(x_0\right) \\
&\hspace{2.5cm} + \left(C^{\partial \partial \mathcal{O}^\prime}_{\mathcal{O}_1 \mathcal{O}_2}\right)^{\mu\nu} \partial_\mu \partial_\nu \mathcal{O}^\prime_g\left(x_0\right)\bigg] + \dots \, .\nonumber
\end{align}
\end{subequations}
The coefficient of $\mathcal{O}_g^\prime\left(x_0\right)$ is the curved space OPE coefficient: 
\begin{equation}
\hat C^{\mathcal{O}'}_{\mathcal{O}_1\mathcal{O}_2}=e^{-\Delta\omega_1-\Delta\omega_2}\Big(C^{\mathcal{O}'}_{\mathcal{O}_1 \mathcal{O}_2}+\Delta'\big(C^{\partial\partial\mathcal{O}'}_{\mathcal{O}_1 \mathcal{O}_2}\big)^{\mu\nu}\partial_\mu\partial_\nu\omega(x_0)\Big).
\label{eq:conf_flat_OPE}
\end{equation}
The first term is what one would naively obtain from rescaling the external operators; the second term comes entirely from Eq.~\eqref{eq:descendant_mixing}.

Using the flat space OPE coefficients in Eq.~\eqref{eq:flatOPE1} and Eq.~\eqref{eq:flatOPE2} and covariantizing, we arrive at 
\begin{equation}
\hat C^{\mathcal{O}^\prime}_{\mathcal{O}_1\mathcal{O}_2}=\frac{\g}{s^{2\Delta-\Delta^\prime}}\Big(D_{12}^{a/(2\Delta_0)}-b\,s^2R\Big),
\label{eq:conf_flat_OPE_final}
\end{equation}
with
\begin{subequations}
    \begin{align}
        a &= \Delta + \frac{\Delta^\prime \left(\Delta^\prime - 2\right)}{4\left(\Delta^\prime + 1\right)},\\
        b &= \frac{a}{12\Delta_0} + \frac{\left(\Delta^\prime\right)^2}{16\left(\Delta^\prime + 1\right)\left(\Delta_0 - \Delta^\prime\right)}.
    \end{align}
\end{subequations}
This matches the result obtained in the main text, Eq.~\eqref{eq:conformal_3pt_final}. The method presented here, based on a conformally flat background, agrees with Ref.~\cite{Das:2026tjz}. 


\bibliographystyle{JHEP}
\bibliography{biblio.bib}

\end{document}